\documentclass[a4paper,11pt]{article}
\pdfoutput=1 % if your are submitting a pdflatex (i.e. if you have
\usepackage{jcappub}
\usepackage[T1]{fontenc} % if needed
\usepackage{comment}
\usepackage{xcolor,cancel}
\usepackage{orcidlink}
\usepackage{adjustbox}
\usepackage{slashed} 
\usepackage{float} 
\usepackage{physics}
\usepackage{pbox}
\usepackage{subfigure}
\usepackage{array,multirow,makecell}
\usepackage{titlesec}

\newcommand{\eq}[1]{Eq.~(\ref{#1})}
\newcommand{\dis}[1]{\begin{equation}\begin{split}#1\end{split}\end{equation}}
\newcommand{\VEV}[1]{\langle #1 \rangle}

\newcommand{\bfrac}[2]{\left(\frac{#1}{#2} \right)  }

\newcommand\gev{\,{\rm GeV}}
\newcommand\mev{\,{\rm MeV}}

\newcommand\unitev{\,{\rm eV}}

\DeclareUnicodeCharacter{2212}{-}

\graphicspath{{./figures/}}% Put all figures in this directory.

\title{Reconciling Cosmological Tensions with Inelastic Dark Matter and Dark Radiation in a \texorpdfstring{$\boldsymbol{U(1)_D}$}{U(1)_D} Framework}

\author[a]{Wonsub Cho\,\orcidlink{0000-0003-2493-4628}}
\emailAdd{sub526@skku.edu}		
    
\author[a,b]{Ki-Young Choi\,\orcidlink{0000-0001-7604-6644}}	
\emailAdd{kiyoungchoi@skku.edu}

\author[a]{Satyabrata Mahapatra\,\orcidlink{0000-0002-4000-5071}}
\emailAdd{satyabrata@g.skku.edu}

\affiliation[a]{Department of Physics and Institute of Basic Science, Sungkyunkwan University, Suwon 16419, Korea}
\affiliation[b]{Korea Institute for Advanced Study, Seoul 02455, Korea}

\abstract{We propose a novel and comprehensive particle physics framework that addresses multiple cosmological tensions observed in recent measurements of the Hubble parameter, $S_8$, and Lyman-$\alpha$ forest data. Our model, termed `{\bf SIDR$+\boldsymbol{z_t}$}' (Self Interacting Dark Radiation with transition redshift), is based on an inelastic dark matter (IDM) scenario coupled with dark radiation, governed by a $U(1)_D$ gauge symmetry. This framework naturally incorporates cold dark matter (DM), strongly interacting dark radiation (SIDR), and the interactions between these components. 
The fluid-like behavior of the dark radiation component which originates from the self-quartic coupling of the $U(1)_D$ breaking scalar can suppress the free-streaming effects.
Simultaneously, the interacting DM-DR system can attenuate the matter power spectrum at small scales.
The inelastic nature of DM provides a distinct temperature dependence for the DM-DR interaction rate determined by the mass-splitting between the inelastic dark fermions which is crucial for resolving the Ly-$\alpha$ discrepancies. We present a cosmologically consistent analysis of the model by solving the relevant Boltzmann equations to obtain the energy density and number density evolution of different species of the model. The DR undergoes two ``steps" of increased energy density when the heavier dark species freeze out and become non-relativistic, transferring their entropy to the dark radiation and enhancing $\Delta N_{\rm eff}$. The analysis showcases the model's potential to uphold the Big Bang Nucleosynthesis (BBN) prediction of $\Delta N_{\rm eff}$ but dominantly producing additional contributions prior to recombination, while simultaneously achieving correct relic density of DM though an hybrid of freeze-in and non-thermal production.}

\begin{document}

\maketitle

%--------------------------------------------------------------------------------------------------
%%%%%%%%%%%%%%%%%%%%%%%%%%%%%%%%%%%%%%%%%%%%%%%%%%%%%%%%%%%%%%%%%%%%%%%%%%%%%%%%%%%%%%%%%%%%%%%%%%%

\section{INTRODUCTION}\label{intro}

The standard model of cosmology, $\rm \Lambda CDM$, has been remarkably successful in explaining a wide range of observational data, from the cosmic microwave background (CMB) to large-scale structure (LSS) formation. However, recent high-precision measurements have revealed tensions that challenge this paradigm. Among these discrepancies, the Hubble tension and the $S_8$ tension stand out as particularly significant~\cite{Planck:2018vyg,Abdalla:2022yfr,DiValentino:2021izs}. The Hubble tension refers to the discrepancy between the locally measured Hubble constant ($H_0$) using Supernovae Ia (SNe Ia) calibrated by Cepheids~\cite{Riess:2021jrx}, and the value inferred from the CMB observations by the Planck satellite~\cite{Planck:2018vyg}. On the other hand, the $S_8$ tension involves a mismatch between the amplitude of matter fluctuations inferred from LSS surveys~\cite{DES:2021wwk,Philcox:2021kcw,KiDS:2021opn} and the CMB data~\cite{Planck:2018vyg,ACT:2020gnv}. These tensions have reached levels of 5$\sigma$ and 3$\sigma$, respectively. Additionally, there is a significant tension of approximately 4.9$\sigma$ between the linear matter power spectrum (MPS) inferred from combined Planck CMB, baryon acoustic oscillations (BAO), and supernovae data, and that from the eBOSS Ly-$\alpha$ forest observations~\cite{Rogers:2023upm}. These discrepancies  indicate potential inadequacies in the standard cosmological model and suggest the need for novel scenarios to tackle these issues.
 
Many mechanisms have been proposed to ease the $H_0$ tension, most of which involve the presence of additional energy around the time of matter-radiation equality such that it reduces the sound horizon during the era leading up to recombination~\cite{Bernal:2016gxb,Knox:2019rjx,Aylor:2018drw, Evslin:2017qdn, Poulin:2018cxd,Vagnozzi:2023nrq}. Such models are usually referred to as ``early-time solutions" (for example see \cite{Bringmann:2018jpr, Pandey:2019plg,Agrawal:2019lmo,Aboubrahim:2022gjb,Verde:2023lmm,Freedman:2023jcz,Vagnozzi:2021gjh}). One such proposed solution involves additional relativistic degrees of freedom, often parameterized by the effective number of neutrino species $N_{\rm eff}$~\cite{Aloni:2021eaq, Aboubrahim:2022gjb,  Brust:2017nmv, Ko:2016uft,Escudero:2021rfi,Ko:2017uyb,Vagnozzi:2019ezj}. Free-streaming dark radiation has been considered as a potential remedy for the Hubble tension by contributing to the early-time expansion rate. However, such models tend to exacerbate the $S_8$ tension since
they require a larger value of $\Omega_m$ to maintain the redshift at matter-radiation equality fixed, which increases the value of $\sigma_8$~\cite{DiValentino:2020vvd,Nunes:2021ipq, Joseph:2022jsf}. Free-streaming dark radiation models face significant challenges in fitting CMB polarization data effectively. The modification of the Silk damping tail in these scenarios leads to a poor agreement with observations~\cite{Hou:2011ec}, which is only partially compensated by the inclusion of dark radiation self-interactions. Without self-interactions, free-streaming DR alone is insufficient to resolve the tensions in the CMB polarization spectrum.
Such scenarios also worsen the fit to Ly-$\alpha$ forest observations, which probe the matter power spectrum at small scales~\cite{Bagherian:2024obh}.
Thus, free-streaming dark radiation alone is insufficient to resolve these cosmological tensions simultaneously.
Recent studies have highlighted the potential of SIDR in alleviating the $H_0$ and $S_8$ tension simultaneously as the correlation between these two quantities is lessened in models with a non-free-streaming radiation~\cite{Blinov:2020hmc,Brinckmann:2022ajr,Ghosh:2021axu,Schoneberg:2023rnx}. Self-interactions suppress free-streaming effects and modify the evolution of cosmological perturbations, leading to a more consistent fit with observational data. Additionally, DM-DR interactions have been proposed as a promising avenue for resolving small-scale structure issues and improving consistency with Ly-$\alpha$ data~\cite{Bagherian:2024obh,Garny:2018byk,Archidiacono:2019wdp}.

Models of SIDR that interact with DM, known as ‘SIDR+' models, have demonstrated the ability to provide a smooth transition in the matter power spectrum and address both the Hubble tension and the $S_8$ tension~\cite{Buen-Abad:2015ova,Lesgourgues:2015wza,Pan:2018zha}. Additionally, the Wess Zumino Dark Radiation (WZDR) model with DM-DR interactions known as ‘WZDR+', or the stepped partially acoustic dark matter (SPartAcous) model, further refines these interactions by introducing a break in the MPS before recombination at a specific redshift $z_t$~\cite{Buen-Abad:2023uva,Joseph:2022jsf,Schoneberg:2023rnx}. These models have been shown to be consistent with Ly-$\alpha$ constraints, as highlighted in~\cite{Garny:2018byk,Archidiacono:2019wdp,Bagherian:2024obh} ensuring that the proposed interactions do not conflict with the observed small-scale structure of the universe. This comprehensive approach of integrating SIDR with DM interactions offers a promising pathway to resolve multiple cosmological tensions simultaneously.

In this paper, we introduce a novel and cosmologically consistent particle physics framework designed to resolve these  puzzles by IDM and SIDR within a $U(1)_D$ gauge symmetry extension of the Standard Model (SM) which we refer to as {\bf `SIDR+$\boldsymbol{z_t}$'}. The self-interactions of the dark radiation component result in fluid-like dynamics, effectively diminishing free-streaming and anisotropic stress, a crucial factor for addressing the $S_8$ tension.
%Additionally, we demonstrate that our model can naturally generate a matter power spectrum that aligns with Ly-$\alpha$ observations. 
Additionally, our model may  produce a matter power spectrum that might be consistent with Lyman-$\alpha$ observations.
The interplay between the SIDR and its interactions with the IDM provides a framework to reconcile Ly-$\alpha$ constraints with the observed characteristics of small-scale structures. Our ‘SIDR+${z_t}$' model facilitates momentum transfer between the DM and DR through the mediation of a heavier dark fermion. This scattering rate displays a unique temperature dependence that is influenced by the mass-splitting of the IDM fermions, setting it apart from the ‘SIDR+' and ‘WZDR+' scenarios. Furthermore, the matter power spectrum exhibits a smooth suppression corresponding to a redshift $z_t$, which is determined by the mass-splitting of IDM fermions. Thus, our scenario presents a more generalized approach compared to the ‘SIDR+' model, featuring a transition redshift for the suppression of the matter power spectrum.
As the dark sector is realized through an abelian $U(1)_D$ gauge symmetry, dark sector particles are produced via the freeze-in mechanism and non-thermal contributions~\cite{Baer:2014eja}. This approach allows for a significant contribution to $N_{\rm eff}$ from SIDR without violating BBN constraints on $N_{\rm eff}$, while DM achieves a correct relic abundance simultaneously. In our setup, DR also undergoes two “steps" increase in its energy density, when the heavier dark species (DM and the $U(1)_D$ gauge boson) freeze out and become non-relativistic, thereby transferring their entropy to the DR~\cite{Aloni:2021eaq,Joseph:2022jsf}. Consequently it increases the overall energy density of DR and thus helps in achieving an enhanced $\Delta N_{\rm eff}$.

The rest of the manuscript is built up as follows: we begin by briefly discussing the concept of this setup and various observational constraints in section~\ref{sec:concept} and explain the details of the model in section~\ref{sec:model} along with its cosmological phenomenology that solves the cosmological tensions in section~\ref{pheno}. Subsequently in section~\ref{sec:production} we discuss the production of DM as well as the generation of dark radiation by solving the set of coupled Boltzmann equations and present the result of our numerical calculations. Finally we conclude the manuscript in section~\ref{sec:conclude}. 

%\newpage
\section{Observational constraints and the interacting DM-DR}\label{sec:concept}

Assuming the $\Lambda$CDM model, the Planck CMB data predicts a present Hubble constant value $H_0=67.27\pm0.6 ~{\rm km/s/Mpc}$ at $68\%$ C.L.~\cite{Planck:2018vyg} ($H(t) \equiv a^{-1}~{da}/{dt}$ and $a_0a^{-1} = 1 + z$ with scale factor $a$, and subscript denotes the value at present), which is based on the direct measurement of the angular size of the acoustic scale in the CMB power spectrum. In contrast, direct measurements in the local universe which determine $H_0$ by constructing a distance-redshift relation known as the cosmic ``distance ladder," show a discrepancy with this result. For example, the latest measurement from the SH0ES collaboration reports $H_0=73.04\pm1.04 ~{\rm km/s/Mpc}$~\cite{Riess:2021jrx}. There are actually two sets of measurements rather than just these two values. All the indirect, model-dependent early estimates, such as those from the CMB and BAO experiments, agree among themselves, while all the direct, late-time {$\Lambda$CDM-independent }measurements, such as those from distance ladders and strong lensing, also agree among themselves~\cite{Abdalla:2022yfr,DiValentino:2021izs}.

The $S_8$ parameter, defined as \(S_8 = \sigma_8 \left( \frac{\Omega_m}{0.3} \right)^{1/2}\) (where \(\Omega_m \equiv \rho_m / \rho_c\) is the matter density today and \(\sigma_8\) is the root mean square amplitude of linear matter density fluctuations), characterizes the amplitude of matter density fluctuations in the Universe.  CMB experiments typically yield higher \(S_8\) values, for example, the primary CMB measurement by Planck~\cite{Planck:2018vyg} reports \(S_8 = 0.834 \pm 0.016\). In comparison, weak lensing surveys such as KiDS reports \(S_8 = 0.759 \pm 0.024\)~\cite{KiDS:2020suj} and DES reports \(S_8 = 0.776 \pm 0.017\)~\cite{DES:2021wwk}, suggesting the universe is less “clumpy” than predicted by the standard \(\Lambda\)CDM model. While this tension is somewhat less pronounced, its consistency across several data sets is noteworthy.

Numerous models have been proposed to address the Hubble tension. These include increasing the effective number of relativistic species  $N_{\rm eff}$~\cite{Aloni:2021eaq,Aboubrahim:2022gjb,  Brust:2017nmv, Ko:2016uft,Escudero:2021rfi,Ko:2017uyb,Papanikolaou:2023oxq}, incorporating neutrinos with significant self-interactions~\cite{Gelmini:2019deq,Seto:2021tad,Huang:2021dba,Arias-Aragon:2020qip} or interactions with dark matter~\cite{Buen-Abad:2015ova,Raveri:2017jto,Lesgourgues:2015wza,Choi:2020pyy,Chacko:2016kgg,Cyr-Racine:2015ihg}, considering decaying dark matter~\cite{Pandey:2019plg,Blinov:2020uvz,Choi:2020tqp,Fuss:2022zyt,Aboubrahim:2022gjb,Cheek:2024fyc,Fuss:2024dam,Cheek:2022yof}, or introducing early dark energy~\cite{Seto:2021xua,Agrawal:2019lmo,Freese:2021rjq}. For recent reviews one may refer to~\cite{Schoneberg:2021qvd,Abdalla:2022yfr,DiValentino:2021izs,Kamionkowski:2022pkx,Hu:2023jqc} and references therein. 

Among the various scenarios mentioned, models incorporating self interacting dark radiation and its interactions with dark matter (‘SIDR+' model) have shown promise in addressing multiple cosmological puzzles simultaneously, as discussed in the previous section. Here, we review the observational constraints on such models and their effectiveness in resolving these tensions.  Additionally, we discuss how these models fare against constraints from Ly$\alpha$ forest data and other LSS observations. 

The ‘SIDR+' model introduces two additional key parameters: $\Delta N_{\rm eff}$, representing the contribution of dark radiation, and $R_\Gamma \equiv \Gamma/H$, which characterizes the interaction rate between DM and DR relative to Hubble expansion.  $\Delta N_{\rm eff}$ is defined as the difference between the effective number of relativistic species, $N_{\rm eff}$, and the standard model prediction, $N^{\rm SM}_{\rm eff}$:
\dis{
\Delta N_{\rm eff} = N_{\rm eff} - N^{\rm SM}_{\rm eff},\quad{\rm with}\quad N_{\rm eff} = \frac{\rho_{r} - \rho_\gamma}{\rho_{\nu_L}}\,,
}
where $\rho_\gamma = \frac{\pi^2}{15} T_\gamma^4$ is the energy density of photons, and $\rho_{\nu_L} = \frac{7}{8} \left( \frac{T_\nu}{T_\gamma} \right)^4 \rho_\gamma$ corresponds to the energy density of a single active neutrino
species.
In the absence of additional radiation, the standard model accurately predicts $N^{\rm SM}_{\rm eff} = 3.046$, which arises from the entropy conservation argument following electron-positron annihilation, where $T_\nu/T_\gamma = (4/11)^{1/3}$. Consequently, $\Delta N_{\rm eff}$ can be expressed as:
\dis{
\Delta N_{\rm eff} = \frac{\rho_{\rm DR}}{\rho_{\nu_L}} = \frac{8}{7} \left( \frac{T_\nu}{T_\gamma} \right)^{-4} \frac{\rho_{\rm DR}}{\rho_\gamma} \simeq \frac{8}{7} \left( \frac{11}{4} \right)^{4/3} \frac{\rho_{\rm DR}}{\rho_\gamma},
\label{DNeff2}
}
where the last equation is applicable after electron-positron annihilation in the instantaneous-decoupling approximation, {\it i.e.} for $T_\gamma \lesssim 0.5 \, \text{MeV}$. It is important to note that models that attempt to resolve the $H_0$ tension by adding extra degrees of freedom must only affect the dynamics during the CMB era and not impact BBN. This is critical due to the precise predictions of light element production during BBN, which place tight constraints on the effective degrees of freedom. Current data shows \(N_{\rm eff}^{\rm BBN} = 2.88^{+0.19}_{-0.37} \) at 68\% CL~\cite{Cyburt:2015mya,Pitrou:2018cgg,Yeh:2022heq,Schoneberg:2024ifp}, closely matching the SM prediction of \(N_{\rm eff}^{\rm SM} \simeq 3.046\)~\cite{Planck:2018vyg, ParticleDataGroup:2024cfk}. Any changes in \(N_{\rm eff}\) during BBN could significantly alter the Universe's expansion rate and light element synthesis. Therefore, models must introduce additional degrees of freedom for the CMB without violating BBN constraints, significantly narrowing the range of viable options.

Recent analyses of the CMB data within the ‘SIDR+' model reveal best-fit values for these parameters: $\Delta N_{\rm eff} \simeq 0.7$ and $R_\Gamma \simeq 0.056$. These values show consistency with the LSS data including Ly-$\alpha$ while significantly reducing the Hubble tension to approximately $2\sigma$~\cite{Blinov:2020hmc,Allali:2024cji,Bagherian:2024obh}.

In this context, the WZDR+ model features three parameters: $N_{\rm IR}$, $R_\Gamma$, and $z_t$, which represent the amount of DR around recombination, the DM-DR interaction rate relative to Hubble expansion, and the transition redshift for momentum transfer, respectively. At the transition redshift, an increase in $\Delta N_{\rm eff}$ occurs, leading to what is known as a stepped SIDR. This increase arises from the growth of DR energy density at redshift $z_t$, driven by the annihilation of additional heavy DR, thereby enhancing the fit to the CMB. This is due to its distinct contributions to the low-$l$ and high-$l$ multipoles in the TT power spectrum~\cite{Aloni:2021eaq}. The model favors parameters of $N_{\rm IR} \simeq 0.59$, $R_\Gamma \simeq 0.07$, and $\log z_t \simeq 4.25$, based on analyses of weak lensing, CMB lensing, full-shape galaxy clustering, and the e-BOSS Ly$\alpha$ data set (collectively referred to as the $\mathcal{DHL}$ data set)~\cite{Bagherian:2024obh}.

In the following section, we introduce a minimal model termed the `{\bf SIDR+$\boldsymbol{z_t}$}' model, which bridges the ‘SIDR+' and ‘WZDR+' models. This model extends the SIDR+ framework by incorporating a free parameter $z_t$, which influences the MPS similarly to the WZDR+ model. The break in the MPS occurs at $k_t$, suppressing modes for $k > k_t$. Our model introduces parameters $\Delta N_{\rm eff}$, $R_\Gamma$ along with $z_t$, which we hypothesize could help address the cosmological tensions. However, further analysis is needed to confirm these effects.
% \cred{Thus, the parameters $\Delta N_{\rm eff}, R_{\Gamma}$ along with $z_t$ effectively alleviate the $H_0$-tension,  align with the $S_8$ values from
% weak lensing and full-shape LSS data while remaining consistent with the Ly-$\alpha$ observations.} 
Also our model features two “stepped" dark radiation {fluids} which causes an increase in $\Delta N_{\rm eff}$ in the era prior to recombination. 
% which might significantly improve the combined fit to CMB, BAO, and SH0ES data~\cite{Aloni:2021eaq,Joseph:2022jsf}.}

%%%%%%%%%%%%%%%%%%%%%%%%%%%%%%%%%%
\section{Minimal Model for  `SIDR\texorpdfstring{$\boldsymbol{+z_t}$}{}'}\label{sec:model}
%%%%%%%%%%%%%%%%%%%%%%%%%%%%%%%%%%
In this section, we propose a minimal framework incorporating inelastic dark matter with an additional $U(1)$ gauge interaction to address the prevailing cosmological challenges.
We consider an extended gauge symmetry $\mathcal{G}$ by an additional $U(1)_D$ gauge interaction to the SM gauge symmetry $\mathcal{G}_{\rm SM}$, {\it i.e.} $\mathcal{G}\equiv\mathcal{G}_{\rm SM}\otimes U(1)_{D}$ featuring two vector like Dirac fermion singlets $\chi$ and $\psi$ and a singlet scalar $\Phi$.  The relevant Lagrangian can be written as:
\begin{eqnarray}
\mathcal{L} &\supset& i \overline{\psi} \gamma^\mu D_\mu \psi -M_{\psi} \overline{\psi}\psi + i \overline{\chi}  \gamma^\mu D_\mu \chi -M_{\chi} \overline{\chi}\chi +(D_\mu \Phi)^\dagger (D^\mu\Phi)- y \overline{\chi} \Phi \psi \nonumber\\&-&\frac{1}{4} F^{\mu\nu}_X {F^{X}}^{\mu\nu} -\frac{\epsilon}{2} F^{\mu\nu}_X B^{\mu\nu}+{\rm h.c.}
\label{Eq:lag}
\end{eqnarray}
Here, $\psi$ and $\chi$ are singlets under the SM gauge group, with masses $M_{\psi}$ and $M_{\chi}$, respectively. Their charges under $U(1)_D$ are $Q_\psi$, $Q_\chi$, and $Q_\phi$, with the condition $Q_\chi = Q_\psi + Q_\phi$ to maintain gauge invariance. We choose $Q_\chi = 1$, $Q_\phi = 1$, and $Q_\psi = 0$. Being vector-like, $\psi$ and $\chi$ do not introduce any triangle anomalies, as all SM fermions are trivially charged under $U(1)_D$. The covariant derivative is given by $D\mu = \partial_\mu - i g_D Q_D (Z_D)_\mu$, where $g_D$ and $Q_D$ are the gauge coupling and dark gauge charge, respectively. The last term in the Lagrangian~\eq{Eq:lag}, represents kinetic mixing between the $U(1)_D$ and $U(1)_Y$ gauge bosons, parameterized by the kinetic mixing parameter $\epsilon$, a free parameter of the theory.

The scalar $\Phi$ being charged under $U(1)_D$ can break the symmetry when it acquires a VEV $\langle \Phi \rangle = v_\phi$ and generate the mass of the corresponding new gauge boson $Z_D$, {\it i.e.} {$M_{Z_{D}}=g_D Q_\phi v_\phi$.}
The scalar potential can be written as
 \begin{equation}
     V(H, \Phi)= -\mu^2_H \left(H^\dagger H \right) + \lambda_H \left(H^\dagger H \right)^2 -\mu^2_{\Phi} \left(\Phi^\dagger \Phi\right) + \lambda_\Phi \left(\Phi^\dagger \Phi \right)^2  + \lambda_{H \Phi} \left(H^\dagger H\right)\left(\Phi^\dagger \Phi\right)\,.
 \end{equation}

The scalar $\Phi$ is parameterized with $\Phi=(\phi+v_\phi+i\eta)/\sqrt{2}$. When $\Phi$ acquires a VEV and breaks the $U(1)_D$ symmetry, it also mixes with the SM Higgs {$H^T=(0, \frac{1}{\sqrt{2}}(v +h))$  }, with $v=246 {\rm GeV}$. 
The mixing parameter $\gamma$ is given by :
 \begin{equation}
	\tan 2\gamma=\frac{\lambda_{H\Phi}v v_\phi}{\lambda_H v^2-\lambda_\Phi v^2_\phi}\,.
\end{equation}  
After mass diagonalization, we get the mass eigen states $h_1$ and $h_2$. We recognise $h_1$ as the SM Higgs and $h_2$ as the second Higgs.
 We work in the limit $\lambda_{H\Phi}\to 0$ {(or $\gamma\rightarrow 0$)}, and thus only kinetic mixing portal is relevant for freeze-in mechanism. {In this limit, $h_2$ is identified as $\phi$.} It also isolates $\phi$ from the SM sector. The scalar $\phi$ will be identified as DR with its mass much smaller than $\unitev$ achieved by tuning the coupling $\lambda_\Phi$.

After $U(1)_D$ breaking, the fermions $\psi$ and $\chi$ mix because of the Yukawa coupling among them as $\Phi$ acquires a VEV. In the basis $\left(\chi,\psi\right)^T$, the mass matrix for the fermions can be written as
\begin{equation}
 \left(\begin{array}{cc}
M_\chi &y v_\phi/\sqrt{2} \\
y v_\phi/\sqrt{2} & M_\psi \\
\end{array}
\right)\,.   
\end{equation}
After the diagonalisation of the above mass matrix we get the physical states
\begin{equation}
    \xi_1 =\cos\beta ~\chi -\sin\beta~ \psi ~~~~~~~{\rm and}~~~~~~
\xi_2=\sin\beta~ \chi + \cos\beta~ \psi,
\end{equation}
with two mass eigenvalues
\dis{
M_{\xi_{1,2}} = \frac12\left( M_\chi+M_\psi \pm \sqrt{(M_\chi-M_\psi)^2 +2 y^2v_\phi^2} \right),
}
and the mixing parameter
\begin{equation}
\tan 2\beta=\frac{\sqrt{2}yv_\phi}{M_\psi-M_\chi}.
\end{equation}
From the inverse transformation, we can obtain a relation
\begin{equation}
    \delta\equiv M_{\xi_2}-M_{\xi_1} = \frac{\sqrt{2}yv_\phi}{\sin2\beta}.
\end{equation}
Thus in the physical basis, the interactions in the dark sector can be written as
\begin{eqnarray}
\mathcal{L} &\supset& ig_D(Z_D)_\mu\left[c^2_\beta~~ \overline{\xi_1} \gamma^\mu \xi_1 + s^2_\beta~~ \overline{\xi_2} \gamma^\mu \xi_2 +c_\beta s_\beta~(\overline{\xi_1} \gamma^\mu \xi_2+\overline{\xi_2} \gamma^\mu \xi_1)\right] \nonumber\\
&-& y (\cos\gamma~ h_2 + \sin\gamma ~h_1) \left[c_{2\beta} (\overline{\xi_1} \xi_2+\overline{\xi_2} \xi_1) +s_{2\beta} ~(\overline{\xi_2} \xi_2 -\overline{\xi_1} \xi_1)\right],
\label{Lag_phy}
\end{eqnarray}
where $c_{\beta}=\cos\beta$, and $s_{\beta}=\sin\beta$ respectively and we will take $\gamma=0$ identifying $h_2=\phi$ afterwards.
Since the kinetic mixing exists, the relevant interactions between the visible sector and dark sector can be written as:
\begin{equation}
    \mathcal{L} \supset  \epsilon g (Z_D)_\mu \overline{f} \gamma^\mu f + \epsilon g_X \frac{s_{\theta_W}}{c_{\theta_W}}Z_\mu \left[c^2_\beta~~ \overline{\xi_1} \gamma^\mu \xi_1 + s^2_\beta~~ \overline{\xi_2} \gamma^\mu \xi_2 +c_\beta s_\beta~(\overline{\xi_1} \gamma^\mu \xi_2+\overline{\xi_2} \gamma^\mu \xi_1)\right] 
\end{equation}
where $f$ represents standard model fermions, $\theta_W$ is the Weinberg mixing angle, and $g$ is the electro-weak gauge coupling. 
We also consider a gauge singlet scalar $S$ to be the source of a non-thermal contribution to the dark matter and dark radiation. The relevant lagrangian for $S$ can be written as
\begin{equation}
    \mathcal{L} \supset \kappa_1 S \bar{\xi_1} \xi_1 + \kappa_2 S \bar{\xi_2} \xi_2,
\end{equation}
where $\kappa_i$ is Yukawa coupling between $\xi_i$ and $S$ and the corresponding relevant terms in the scalar potential can be written as :
\begin{equation}
    V(S)\supset \mu^2_S S^2 + \lambda_S S^4 + \lambda_{SH} S^2 (H^\dagger H) + \lambda_{S\Phi} S^2 (\Phi^\dagger \Phi) ,
\end{equation}
%\cred{added '-' in front of $\mu_S^2$.}
This scalar $S$ is produced from the interaction $ H H \to SS$ and decays to DM at late time after BBN.

\section{Cosmological phenomenology}
\label{pheno}
In this section, we explore the phenomenological implications of our model, as introduced in Sec.~\ref{sec:model}, with a focus on addressing the cosmological tensions. Our model exhibits several key properties as we will show below: 1) the decay of particle $S$ results in the production of DM and DR, yielding the observed values of $\Omega_{\rm DM} h^2$ and required $\Delta N_{\rm eff}$ after BBN, 2) the DR is characterized with self-interactions, facilitated by its self quartic coupling $\lambda_\phi$, 3) there are interactions between DM and DR, leading to momentum transfer at a rate defined by $R_\Gamma \equiv \Gamma/H$ where $\Gamma$ is the interaction rate and $H$ is the Hubble parameter, 4) the suppression of these interactions occurs at a redshift $z_t$, which is determined by the mass difference $\delta$ between the particles $\xi_1$ and $\xi_2$, and 5) a step-like increase of $\Delta N_{\rm eff}$ during the annihilation of heavy particles $\xi$ and $Z_D$. 
These properties of our model are designed to potentially address discrepancies in $H_0$ and $S_8$ while aiming for consistency with LSS and Ly-$\alpha$ observations. However, further detailed analysis is needed to confirm these effects.
%\cred{As previously discussed, these properties collectively help reconcile the discrepancies in $H_0$ and $S_8$ while remaining consistent with LSS and Ly-$\alpha$.}

\subsection{Production and decay of \texorpdfstring{$\boldsymbol{S}$}{}}
$S$ particles can be produced from the interactions with the Higgs field in the visible sector through interactions like $HH \rightarrow SS$, $W^+W^- \rightarrow SS$ and $H \rightarrow SS$ decay.
Depending on the coupling strength $\lambda_{SH}$,
the abundance of $S$ can be determined by the freeze-out or freeze-in mechanism.

Regardless of the production mechanism, once produced, the $S$ particle becomes non-relativistic at temperatures $T \lesssim M_S$. Consequently, the energy density $\rho_S = M_S n_S$ increases relative to the background radiation. Given that most of this energy is subsequently transferred into DR via the annihilation processes of $\xi_{1,2}$ and $Z_D$, we can approximate $\rho_{\rm DR}$ around the recombination epoch from the $\rho_{S}$ before $S$ decays. Thus with appropriate choice of $\lambda_{SH}$, we can obtain the required $\Delta N_{\rm eff}$ at late time after its decay. 

The $S$ particle primarily decays into $\xi_1$ and $\xi_2$ through Yukawa couplings $\kappa_1$ and $\kappa_2$, respectively. The lifetime of $S$ is given by:
\begin{equation}
\tau_S = \left[ \sum_{i=1,2}\frac{\kappa_i^2 M_S}{8\pi}\right]^{-1} \simeq 16~\text{s} \left(\frac{10^{-12}}{\kappa}\right)^2 \left(\frac{1~\text{GeV}}{M_S}\right),
\end{equation}
where $\kappa^2 \equiv \sum_i \kappa_i^2$.
To ensure that $S$ decays after BBN has completed and that its energy density during BBN is negligible, we impose the condition $\tau_S \gtrsim 100 ~{\rm s}$. This approach is similar to other models, where DR originates from the decay of heavy particles~\cite{Ichikawa:2007jv,Fischler:2010xz,Hooper:2011aj,Bjaelde:2012wi,Choi:2012zna,Hasenkamp:2012ii,Sobotka:2023bzr}.

\begin{figure}[tbp]
    \centering
    \includegraphics[width=0.35\textwidth,height=0.2\textwidth]{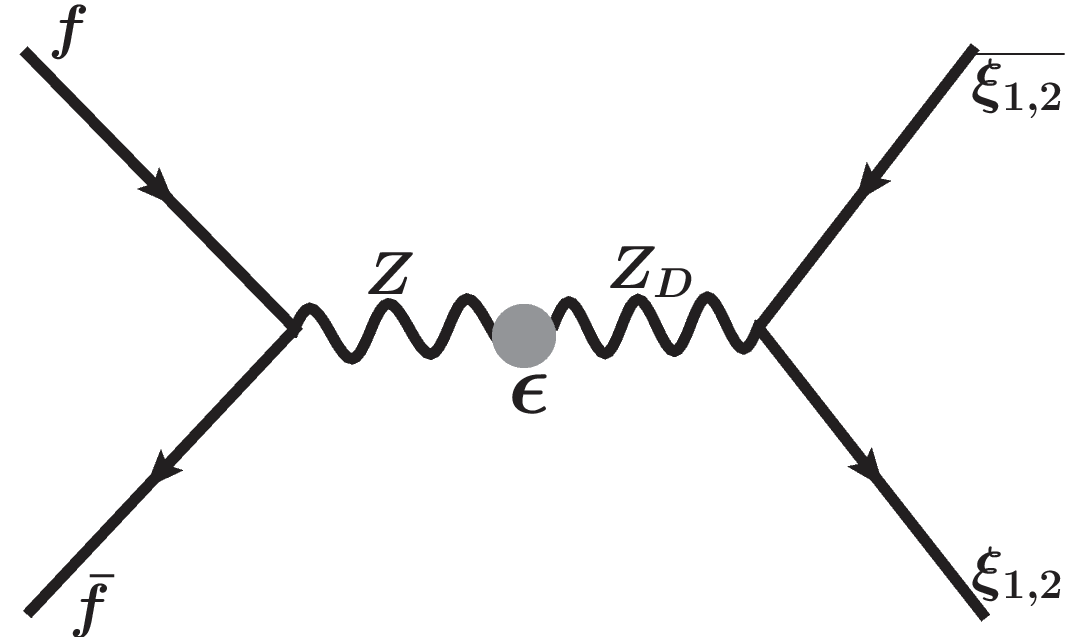}
    \includegraphics[width=0.3\textwidth,height=0.2\textwidth]{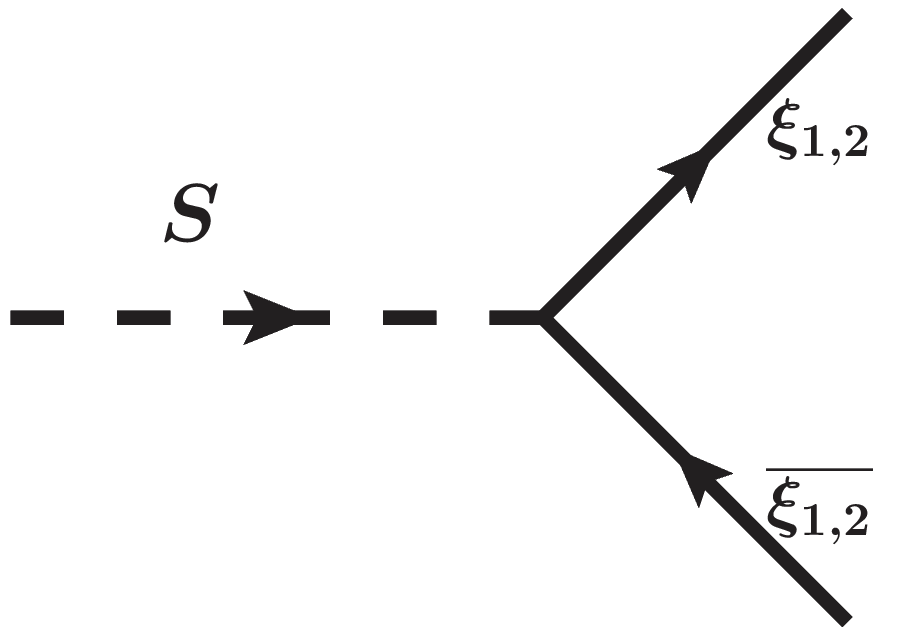}~~~~~
     
    \caption{Production of $\xi_{i}$ via freeze-in from visible sector (left) and from $S$-decay (right).
    }
    \label{fig:S-decay_freezein}
\end{figure}

\subsection{Production of DM \texorpdfstring{$\boldsymbol{\xi_1}$}{}}

In our model, the light fermion $\xi_1$ serves as the DM candidate. It is primarily produced through the decay of $S$, followed by annihilation into $Z_D$ and $\phi$ in the hidden sector. Thus the mass of $\xi_1$ and the associated gauge coupling are determined by the observed DM relic density, as we will demonstrate below.
While $\xi_1$ can also be produced via freeze-in from the visible sector through gauge kinetic mixing, this production mechanism is insufficient to account for the observed DM relic density because of stringent bounds on kinetic mixing parameter from astrophysical and experimental constraints on light gauge bosons with masses of $\mathcal{O}(\rm eV)$~\cite{Caputo:2021eaa}. Fig.~\ref{fig:S-decay_freezein} illustrates the production mechanisms of $\xi_{1,2}$. The left panel depicts the freeze-in production from the visible sector, while the right panel shows the production through $S$-decay.  

Thus the relic density of $\xi_1$ depends on the thermal averaged cross section of $\xi_1$ annihilation into the gauge bosons $Z_D$ as~\cite{ParticleDataGroup:2024cfk}:
\dis{
\Omega_{\xi_1} h^2 \simeq 
\frac{2.6\times 10^{-9}\gev^{-2}}{\VEV{\sigma_{\xi_1} v}}.
}
where the cross-section for the dominant $t$-channel annihilation process is given by~\cite{Pospelov:2008jd}
{\dis{\VEV{\sigma_{\xi_1} v}=\frac{\pi  \alpha_D^2}{M^2_{\xi_1}} \left(1-\frac{M^2_{Z_D}}{M^2_{\xi_1}}\right)^{1/2}.}} 
Here it is worth mentioning that, the other annihilation channel $\xi_1\xi_1 \to \phi \phi$, remains suppressed as compared to the annihilation into gauge bosons. 
Thus to achieve the required DM relic density $\Omega_{\rm DM} h^2 = 0.1200\pm0.0012$~\cite{Planck:2018vyg}, $\alpha_D$ and $M_{\xi_1}$ have to be tuned appropriately. 

The heavier dark fermion $\xi_2$ can decay into $\xi_1$ and $\phi$ due to the Yukawa interaction given in \eq{Lag_phy} and the decay width can be evaluated as~\cite{Borah:2021pet}
\begin{eqnarray}
    \Gamma_{\xi_2\to\xi_1 \phi} &=&  \bfrac{ \cos^2 2\beta \cos^2\gamma}{64 \pi} \bfrac{2y^2}{M^3_{\xi_2}} \left((M_{\xi_2}+M_{\xi_1})^2-M^2_\phi\right)\nonumber \\
    &\times&\left(M^4_{\xi_2}+M^4_{\xi_1}+M^4_\phi-2M^2_{\xi_2}M^2_{\xi_1}-2M^2_{\xi_2}M^2_{\phi}-2M^2_{\xi_1}M^2_{\phi}\right)^{1/2},\\
&\simeq & \bfrac{ \cos^2 2\beta \cos^2\gamma}{4 \pi} y^2\delta,
\end{eqnarray}
where we have assumed $M_\phi=0$ and $\delta \ll M_{\xi_1}$ in the last equation.
Considering small mixing between the $\xi_2$ and $\xi_1$, i.e. $\cos\beta \to 1$, and $\gamma=0$, the corresponding lifetime of $\xi_2$ is
\dis{
\tau_{\xi_2}=\frac{1}{ \Gamma_{\xi_2\to\xi_1 \phi}}  \simeq 8\times10^{-6} {\rm s} \bfrac{10^{-5}}{y}^2\bfrac{10\,\rm eV}{\delta}.
}
Even though both $\xi_1$ and $\xi_2$ are produced from $S$ decay, $\xi_2$ subsequently decays into $\xi_1$. The resulting $\xi_1$ annihilates into $Z_D$ which ultimately produces the dark radiation $\phi$.

\begin{figure}[tbp]
    \centering
       \includegraphics[width=0.35\textwidth,height=0.2\textwidth]{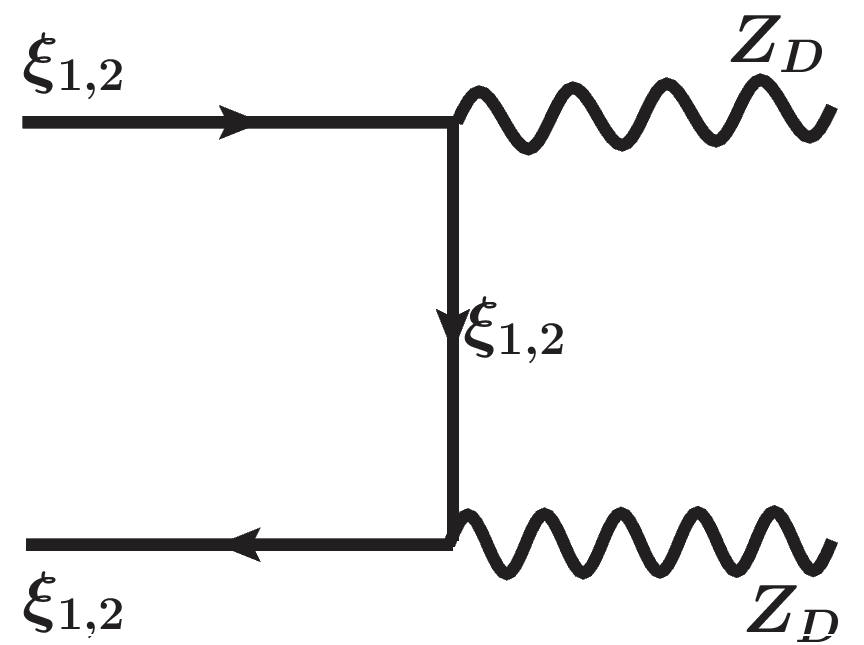}~~~~~
\includegraphics[width=0.32\textwidth,height=0.2\textwidth]{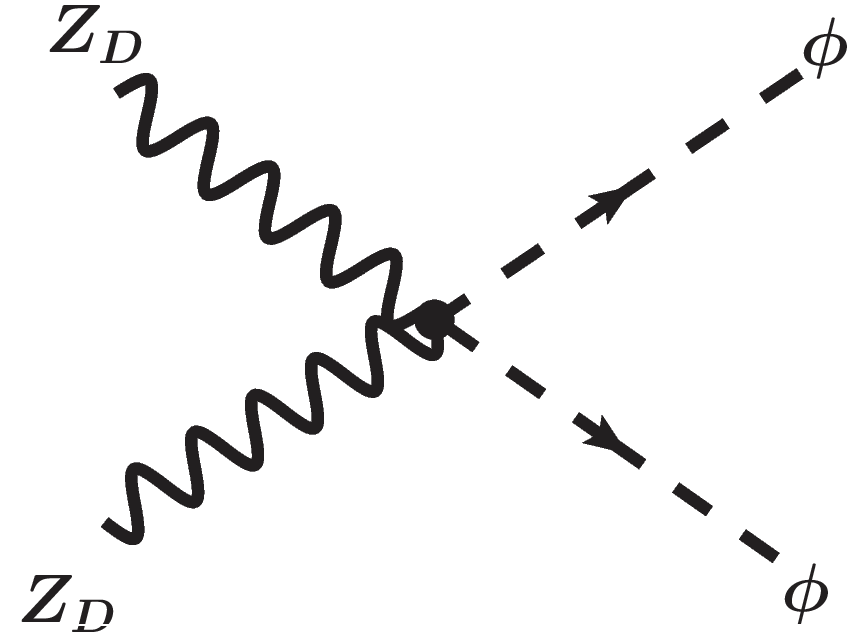}
    \caption{Production of $Z_D$ from annihilation of $\xi_{1,2}$ and its annihilation into $\phi$.}
    \label{fig:annihilation_hiddensector}
\end{figure}

\subsection{Production and annihilation of \texorpdfstring{$\boldsymbol{Z_D}$}{}}
The hidden sector gauge boson $Z_D$ can be produced through two mechanisms: freeze-in from the visible sector and annihilation of $\xi_{1,2}$ particles following $S$ decay. For $M_{Z_D} < m_e$, where $m_e$ is the electron mass, stringent constraints limit the kinetic mixing parameter to $\epsilon \lesssim 10^{-13}$. This severely suppresses $Z_D$ production via freeze-in. However, $Z_D$ production from dark fermion annihilations, following $S$ decay, is sufficient to thermalize both $Z_D$ and $\phi$ in the hidden sector.

$Z_D$ annihilation becomes significant when the hidden sector temperature drops below its mass, $T_h \lesssim M_{Z_D}$. This process results in a small increase in $\Delta N_{\rm eff}$~\cite{Aloni:2021eaq}. Fig.~\ref{fig:annihilation_hiddensector} illustrates the Feynman diagrams for $Z_D$ production through $\xi_{1,2}$ annihilation and its subsequent annihilation into $\phi$. Due to its larger annihilation cross-section, the post-freeze-out abundance of $Z_D$ is significantly lower than that of $\xi_1$. The remaining $Z_D$ population gradually diminishes as it decays into neutrinos via gauge kinetic mixing. The decay width for this process is given by:
\dis{
\Gamma_{Z_D}=\frac{\epsilon^2 g^2 M_{Z_D}}{96\pi \cos^2\theta_W}.
}
Thus the lifetime of $Z_D$ is given by:
\dis{\tau_{Z_D}=  3.65 \times 10^{12}~{\rm s} \bfrac{10^{-13}}{\epsilon}^2\bfrac{10 \unitev}{M_{Z_D}}.
}

\subsection{Production of DR and \texorpdfstring{$\boldsymbol{\Delta N_{\rm eff}}$}{Delta Neff}}
Dark radiation $\phi$ is produced from the annihilation of $Z_D$ and $\xi_i$, which themselves originate from the decay of $S$. Consequently, the final energy density of DR can be estimated approximately from the energy density of $S$ at the time of its decay. When the heavier species freeze out and become non-relativistic, they transfer their entropy to the lighter species, thereby increasing the overall energy density of DR. This increase can be quantified by the conservation of comoving entropy~\cite{Aloni:2021eaq}. 
We can approximate the change of $\Delta N_{\rm eff}$ in three steps as follows:
\begin{itemize}
\item Decay of $S$ into $\xi$, $Z_D$, and $\phi$: Utilizing the instantaneous decay approximation, we express the energy density at the decay time of
\dis{
\left.\rho_S\right|_{\rm decay}=\rho_{\xi} +\rho_{Z_D} + \rho_{\phi},
}
 and at this point, we also have $\left.\rho_S\right|_{\rm decay}=M_S \left.n_S\right|_{\rm decay}$. After the decay of 
$S$, the increase in $\Delta N_{\rm eff}$ is 
\dis{
\Delta N_{\rm eff} \simeq \frac{8}{7} \left( \frac{11}{4} \right)^{4/3} \frac{\rho_{\xi} + \rho_{Z_D} + \rho_\phi}{\rho_\gamma}.
}
where we have utilized Eq.~\eqref{DNeff2}.
\item Annihilation of $\xi$ into DR $\phi$: After the annihilation of $\xi_1$, the hidden sector composed of $\xi_1$, $Z_D$ and $\phi$ becomes a thermal system of $Z_D$ and $\phi$.
During the freeze-out process of $\xi_1$, the hidden sector is a mixture of both massive and massless particles such that its energy density redshifts more slowly than that of the relativistic particles. This contributes to a relative increase in $\Delta N_{\rm eff}$~\cite{Joseph:2022jsf},
\dis{
 \frac{\left.\Delta N_{\rm eff}\right|_{\rm {after}}}{\left.\Delta N_{\rm eff}\right|_{\rm {before}}}= \bfrac{\frac78 g_*^{\xi_1} + g_*^{Z_D} + g_*^\phi}{ g_*^{Z_D} + g_*^\phi}^{1/3}\simeq 1.23.
}
Here $g_*$ denotes the internal degrees of freedom for different species. Specifically, we use $g_*^{\xi_1}=4$, $g_*^{Z_D}=3$, and $g_*^\phi=1$.
\item Annihilation of $Z_D$ into DR $\phi$: Following the annihilation of $\xi_1$, the hidden sector composed of $Z_D$ and $\phi$, becomes a thermal system of $\phi$ only. During the freeze-out process of $Z_D$, the relative increase is given by
\dis{
 \frac{\left.\Delta N_{\rm eff}\right|_{\rm {after}}}{\left.\Delta N_{\rm eff}\right|_{\rm {before}}}= \bfrac{g_*^{Z_D} + g_*^\phi}{ g_*^\phi}^{1/3}\simeq 1.58.
}
\end{itemize}

%We can approximately write $\Delta N_{\rm eff}$ as:
%{\dis{
%\Delta N_{\rm eff} =& \frac{\rho_{\rm DR}}{\rho_{\nu_L}}\simeq \bfrac{\frac78 g_*^{\xi_1} + g_*^{Z_D} + g_*^\phi}{g_*^\phi}^{1/3} \frac{\left.\rho_S\right|_{\rm decay}}{\rho_{\nu_L}},
%}

{Thus DM annihilation results in a $1.23$-fold increase in $\Delta N_{\rm eff}$ and subsequently, $Z_D$ annihilation leads to a further $1.58$-fold increase.
It is important to note that $\xi_2$ is not included in this calculation due to its relatively weaker coupling to $Z_D$, which is a consequence of the mixing angle $\sin\beta$. This weaker interaction prevents $\xi_2$ from achieving thermal equilibrium with the other dark sector particles, thus excluding it from contributing significantly to these entropy transfer processes. This stepwise increase in $\Delta N_{\rm eff}$ illustrates the cascading energy transfer from heavier to lighter species in the dark sector, ultimately augmenting the DR component.}

To achieve $\Delta N_{\rm eff} \simeq 0.6$ which has been proposed to reconcile the Hubble tension in SIDR model~\cite{Bagherian:2024obh} as discussed in Sec.~\ref{sec:concept}, we find that the required ratio of dark radiation to photon energy densities is $\rho_{\rm DR}/\rho_\gamma = 0.136$ from \eq{DNeff2}. Using $\rho_{\rm DR}= \frac{\pi^2}{30}g_hT_h^4$, we determine that the hidden sector temperature $T_h$ would be approximately 0.77 times the visible sector temperature $T_v$ {\it i.e.} $T_h=0.77 ~T_v$ around the recombination epoch. It is important to note that since the production of DR occurs after BBN, it does not impact the abundances of light elements, even with a significant increase in $\Delta N_{\rm eff}$ near the time of recombination.

\subsection{DR self interaction}
%\subsection{Self-interacting dark radiation}

\begin{figure}[tbp]
    \centering
    \includegraphics[width=0.35\textwidth]{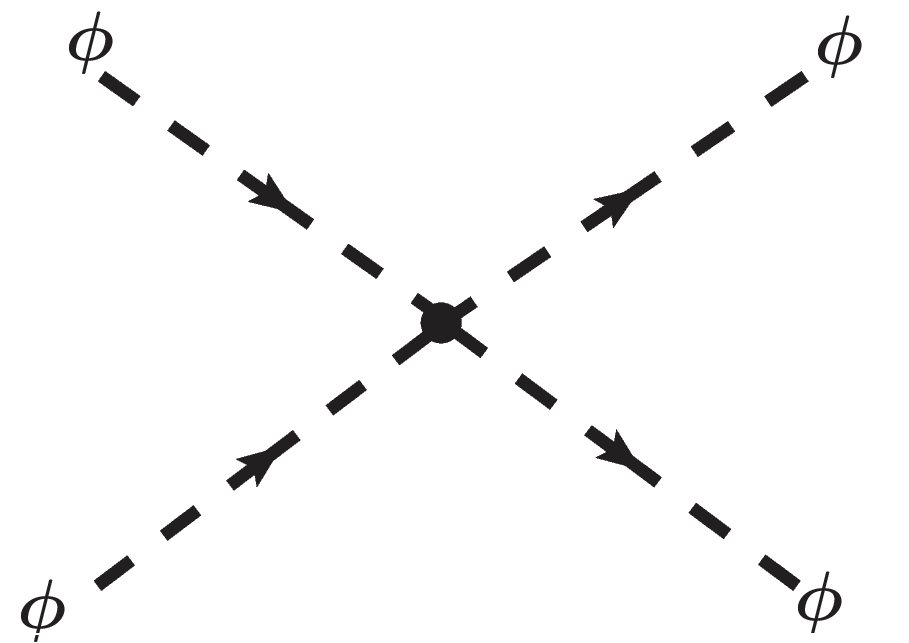}
    \caption{Feynman diagram for quartic coupling of $\phi$ which leads to SIDR.}
    \label{fig:SIDM_diagram}
\end{figure}

The self-interactions among dark radiation can occur through four-point contact interactions, as depicted in Fig.~\ref{fig:SIDM_diagram}, as well as through $H$ or $S$ mediated processes and gauge interactions. These self-interactions prevent the dark radiation from free-streaming, causing it to behave as an ideal relativistic fluid. This behavior has the potential to simultaneously address both the $H_0$ and $S_8$ tensions.
To achieve efficient self-interactions of $\phi$, the coupling strength can be estimated by comparing the interaction rate with the Hubble expansion rate. By requiring the interaction rate $\Gamma_{\phi\phi \leftrightarrow \phi\phi} = n_{\phi} \langle \sigma v \rangle_{\phi \phi} \sim {\frac{9 \zeta(3)}{8\pi^3} \lambda_\phi^2 T_h}$ to be greater than the Hubble rate~\cite{Aboubrahim:2022gjb},
\begin{equation}
   \Gamma_{\phi\phi \leftrightarrow \phi\phi} / H = \frac{\frac{9 \zeta(3)}{8\pi^3} \lambda_\phi^2 \zeta T_v}{\frac{\pi}{3\sqrt{10}}(g_* + \zeta^4)^{1/2} T_v^2}  = 0.13 M_{Pl} \lambda_\phi^2 \frac{\zeta}{(g_* + \zeta^4)^{1/2} T_v} \gtrsim 1
\end{equation}
where $\zeta = T_h/T_v$, and $M_{Pl}$ is reduced Planck mass.
We can estimate the lower limit on the self-interaction coupling at around $T_v \sim 1\unitev$ with $\zeta = 0.77$ and $g_* = 3.36$:
%\cblue{More quantitative details.}
%
\begin{equation}
    \lambda_{\phi} \gtrsim 10^{-13}.
\end{equation}
As discussed in ~\cite{Blinov:2020hmc,Brinckmann:2022ajr}, a non-free-streaming radiation component allows for a larger amount of total energy density in radiation, leading to improvement of the fit to cosmological data compared to models
with only a free-streaming component.

\subsection{DM-DR interaction: \texorpdfstring{$\boldsymbol{\phi+\xi_1 \rightarrow \phi +\xi_1}$}{}}
\begin{figure}[tbp]
    \centering
    \includegraphics[width=0.3\textwidth]{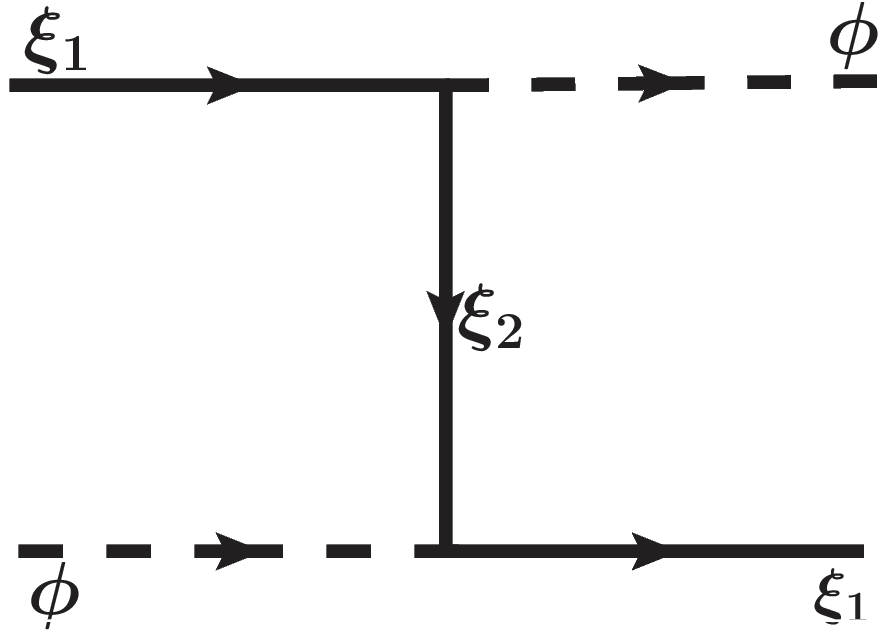}
    ~~~~~\includegraphics[height=0.21\textwidth,width=0.3\textwidth]{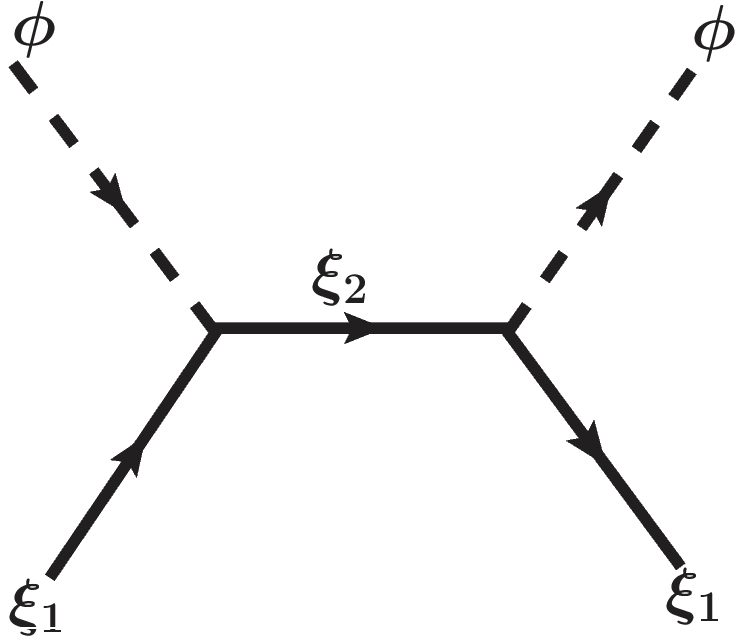}~~~~
    \includegraphics[width=0.3\textwidth]{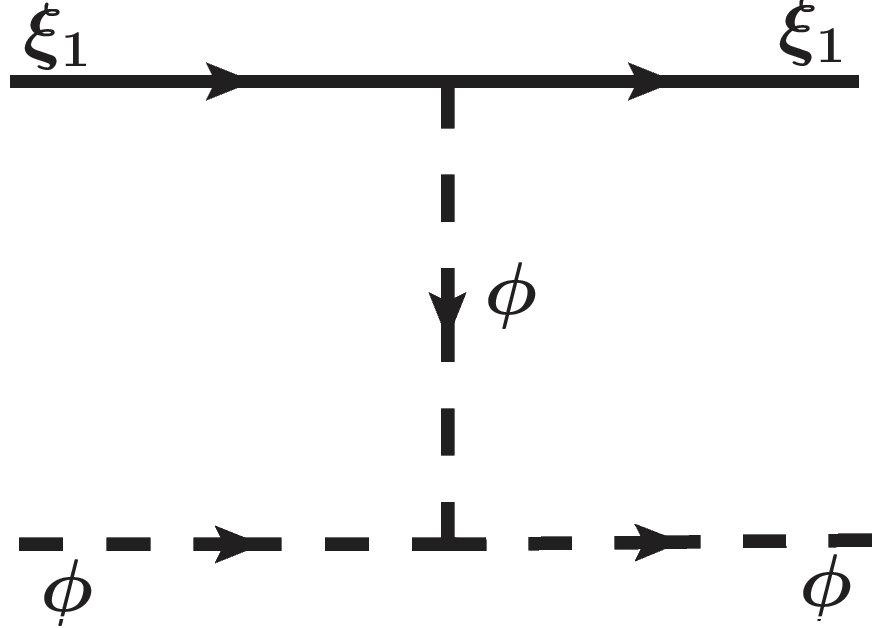}
    \caption{Feynman diagrams for DM-DR scattering.}
    \label{fig:DM-DR_Scattering}
\end{figure}

In our model, the dark fermions $\xi_{1,2}$ interact with the DR $\phi$ via Yukawa interactions.  Thus DM-DR interactions are mediated by $\phi$ as well as $\xi_2$ as shown in Fig.~\ref{fig:DM-DR_Scattering}.
The momentum transfer rate $\Gamma$ is defined as the rate of change of DM momentum due to the friction it experiences as it moves through the DR fluid. Microscopically, this arises from the DM-DR scatterings and to compute it we evaluate the rate of change of DM momentum~\cite{Dvorkin:2013cea}
\dis{
\dot{\vec{p}}_{\rm DM} = -a\Gamma \vec{p}_{\rm DM},
}
where $a$ is the scale factor and the dot denotes the derivative with respect to conformal time. For two-body scattering between DM and DR, $\dot{\vec{p}}_{\rm DM}$ is given by~\cite{Buen-Abad:2015ova,Cyr-Racine:2015ihg}
\dis{
\dot{\vec{p}}_{\rm DM}=\frac{a}{2E_p}\int \frac{d^3 k}{(2\pi)^3 2E_k} \ f(k;T) \int \frac{d^3 k'}{(2\pi)^3 2E_{k'}} \frac{d^3 p'_{\rm DM}}{(2\pi)^3 2E_{P'_{\rm DM}}} &(2\pi)^4 \delta^{(4)} (p_{\rm DM} + k - p'_{\rm DM} -k')\\
&\times\abs{\mathcal{M}}^2 (\vec{p'}_{\rm DM} - \vec{p}_{\rm DM} ).
}
where $k$ is the DR momentum, and {\it prime} denotes the momentum of the outgoing particles. In the limit $M_{\rm DM} \gg k$ and $p_{\rm DM}$, explicitly, after some calculation, the momentum transfer rate $\Gamma$ can be expressed as:
\dis{
\Gamma \simeq \frac{1}{8(2\pi)^3 M_{\rm DM}^3} \int k^3 f(k;T)dk \int d\cos\theta \abs{\mathcal{M}}^2 (1-\cos\theta),
\label{Gam_mt}
}
where $\theta$ is the scattering angle~\footnote{The momentum transfer rate $\Gamma$ can be  approximated as~\cite{Pan:2018zha},
\dis{
\Gamma \simeq \frac{T_{\rm DR}}{M_{\rm DM}}~n_{\rm DR}~\sigma_{\rm mt},
}
where $\sigma_{\rm mt}$ is momentum transfer scattering cross section between DM and DR in the CM frame,
\dis{
\sigma_{\rm mt}=\int d\Omega \frac{d\sigma}{d\Omega} (1-\cos\theta).
}
}.
In our model, DM-DR interaction is dominated by $u$-channel process mediated by $\xi_2$. The $s$-channel process is suppressed by $1/M^4_{\xi_2}$, and the third process (Fig.~\ref{fig:DM-DR_Scattering}) is suppressed due to the small value of $\lambda_\phi$ (as $|\mathcal{M}|\propto y \lambda_\phi v_\phi$).\\

Thus the matrix amplitude squared for the DM-DR scattering is given by
\dis{
\overline{\abs{\mathcal{M}}^2} =& \frac{y^4(1-2\sin^2\beta)^4[(M_\phi^2 +3M_{\xi_1}^2+2M_{\xi_1}M_{\xi_2} -s)^2 +t(s-(2M_{\xi_1}+M_{\xi_2})^2]}{4(u-M_{\xi_2}^2)^2 }\\
\simeq & \frac{y^4(1-2\sin^2\beta)^4 M_{\xi_1}^2}{(k+\delta)^2},
}
where $s$, $t$, and $u$ are the Mandelstam variables with $s+t+u=2M_{\xi_1}^2$ and $s\simeq M_{\xi_1}^2 + 2M_{\xi_1} k$ in the range $M_{\xi_1}^2 \gg \delta$, $t$, and $M_\phi$ is ignored.

In this limit, the momentum transfer rate in \eq{Gam_mt} becomes
\dis{
\Gamma \simeq & ~ \frac{y^4(1-2\sin^2\beta)^4T_h^2}{32\pi^3 M_{\xi_1}}f(x)\\
\simeq &~2.43\times 10^{-34} \gev \bfrac{T_h}{100 \unitev}^2\bfrac{y(1-2\sin^2\beta)}{4.1\times 10^{-6}}^4 \bfrac{0.01\mev}{M_{\xi_1}}, \quad {\rm for} \quad T_h\gg\delta,\\
\simeq &~1.46\times 10^{-43}\gev \bfrac{T_h}{0.1 \unitev}^4\bfrac{y(1-2\sin^2\beta)}{4.1\times 10^{-6}}^4 \bfrac{0.01\mev}{M_{\xi_1}} \bfrac{10\unitev}{\delta}^2, \quad {\rm for} \quad T_h\ll\delta,
}
where  $x=\delta/T_h$ and a function $f(x)$ from the integration is given by
\dis{
f(x) & = x^2(x+3)e^x\Gamma(0,x) -x^2-2x+1,\\
& = 1 \quad {\rm for}\quad  x\ll1,\qquad  6x^{-2}\quad {\rm for}  \quad x\gg1,
}
where $\Gamma(0,x)$ is incomplete Gamma function.
Therefore, the ratio of the momentum transfer rate to the Hubble expansion rate is
\dis{
R_\Gamma\equiv \frac{\Gamma}{H} \simeq 0.07 \bfrac{y(1-2\sin^2\beta)}{4.1\times 10^{-6}}^4\bfrac{0.01\mev}{M_{\xi_1}}, \quad {\rm for} \quad T\gg\delta,\\
}
where we used $T_h=0.77T_v$. 
Clearly, $R_\Gamma$ decreases for smaller temperature $T_h < \delta$.
This implies that the momentum transfer effectively shuts off after crossing the $\delta$ threshold, or below the transition redshift $z_t$, given by $1+z_t \equiv \delta/T_{h_0}$, with $T_{h_0}$ being the present hidden temperature. 
The small mode which enters the horizon at temperature $T_h \gg \delta$ is suppressed, however for  large modes which enter at $T_h \ll \delta$ remain unsuppressed. 
Our SIDR+$z_t$ model extends the SIDR+ framework by introducing a transition redshift $z_t$, which controls the scale at which matter power spectrum suppression occurs. Unlike the standard SIDR+ model, where suppression primarily happens at matter-radiation equality, or the WZDR+ model, where the momentum exchange diminishes after crossing the threshold of mediator mass,  
our model provides a transition in the matter power spectrum, governed by the mass splitting of inelastic dark matter fermions, offering a more flexible mechanism for addressing small-scale structure issues while maintaining consistency with large-scale observations~\cite{Bagherian:2024obh}.
Moreover, our SIDR+$z_t$ model incorporates an additional feature: a two “step" increase in $\Delta N_{\rm eff}$ resulting from the annihilations of $\xi_i$ and $Z_D$ particles. 
%This characteristic potentially enhances the model's fit to CMB data. 
These annihilation events occur at energy scales corresponding to the masses of $\xi_i$ and $Z_D$, and importantly, their effects are distinct from and independent of the suppression scale in the matter power spectrum. This separation of scales provides our model with greater flexibility in addressing multiple cosmological observations simultaneously.

\vspace{1cm}
\noindent

Based on our analytical estimates and a comprehensive scan of the model parameters, we have identified a set of viable values that potentially reconcile the observed cosmological tensions. One such benchmark point is:
\dis{
  M_\xi=0.01 \mev,\quad \delta =3.2 \unitev ,\quad M_{Z_D} =5 \unitev, \quad \alpha_D = 2.8\times 10^{-10}, \quad y=4.1\times 10^{-6}.
 \label{eq:benchmark}
 }
This benchmark point yields the following dependent parameters:
\dis{
v_\phi = 8.4\times 10^{-5}\gev,\quad \lambda_{\phi}= 1.4\times 10^{-12},\quad \sin2\beta=0.153\,.
}
These parameters successfully generate the required DM-DR momentum transfer rate and the correct DM relic density.
Notably, with the specified gauge and Yukawa couplings, the hidden sector particles readily achieve thermal equilibrium. This benchmark point demonstrates the model's capacity to address multiple cosmological issues simultaneously while maintaining consistency with observational constraints.

\section{Boltzmann equations and Numerical results}\label{sec:production}
In this section, we present a comprehensive analysis of the hidden sector dynamics in our inelastic DM and DR model. We begin by outlining the set of coupled Boltzmann equations that govern the evolution of the abundance and energy density of the hidden sector particles. 
By solving these equations numerically, we demonstrate how our model successfully achieves the correct dark matter relic abundance while simultaneously generating the required dark radiation component.
The subsequent discussion will focus on interpreting these results in the context of addressing the cosmological tensions outlined earlier, highlighting the unique features of our ‘SIDR+$z_t$' model.

The hidden and visible sectors in our model are connected through gauge kinetic mixing. To adhere to the stringent constraints on the kinetic mixing parameter for light new abelian gauge bosons, which is essential for reconciling cosmological tensions as demonstrated by our analytical estimates in the previous section, the kinetic mixing parameter must be extremely small. Consequently, the visible and dark sectors never achieve thermal equilibrium with each other.
{In this scenario, hidden sector particles are produced via both freeze-in which involves feeble interactions with the visible sector, and non-thermal production, primarily through the decay of a heavy scalar. This approach provides more flexibility to achieve correct DM relic density and required $\Delta N_{\rm eff}$, enabling us to address multiple cosmological challenges simultaneously} To accurately track the evolution of comoving number densities and energy densities, we solve a set of coupled Boltzmann equations, detailed in Appendix~\ref{apppendix:be}.

\begin{figure}[tbp]
    \centering
    \includegraphics[width=0.48\textwidth]{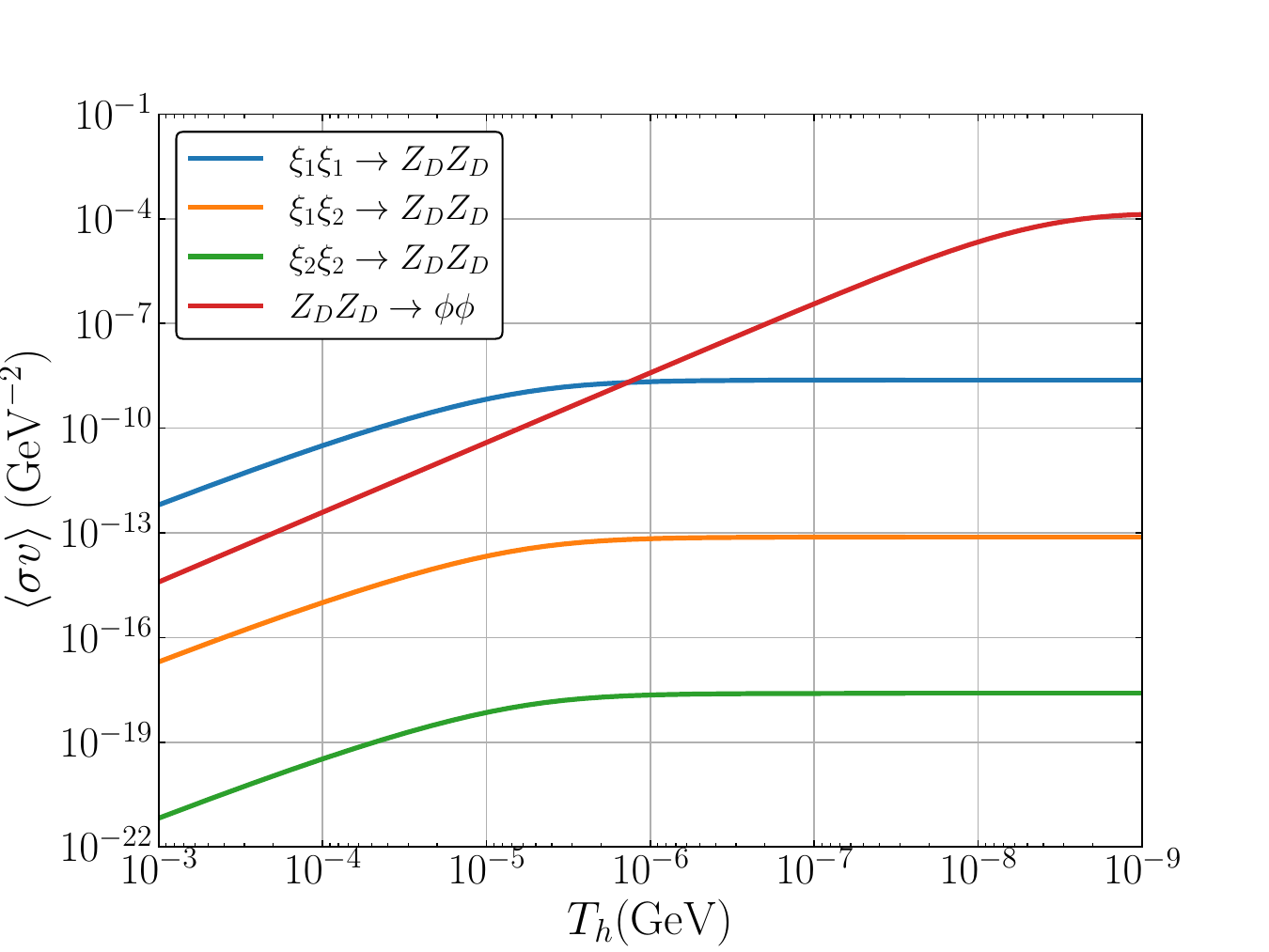}
    \includegraphics[width=0.48\textwidth]{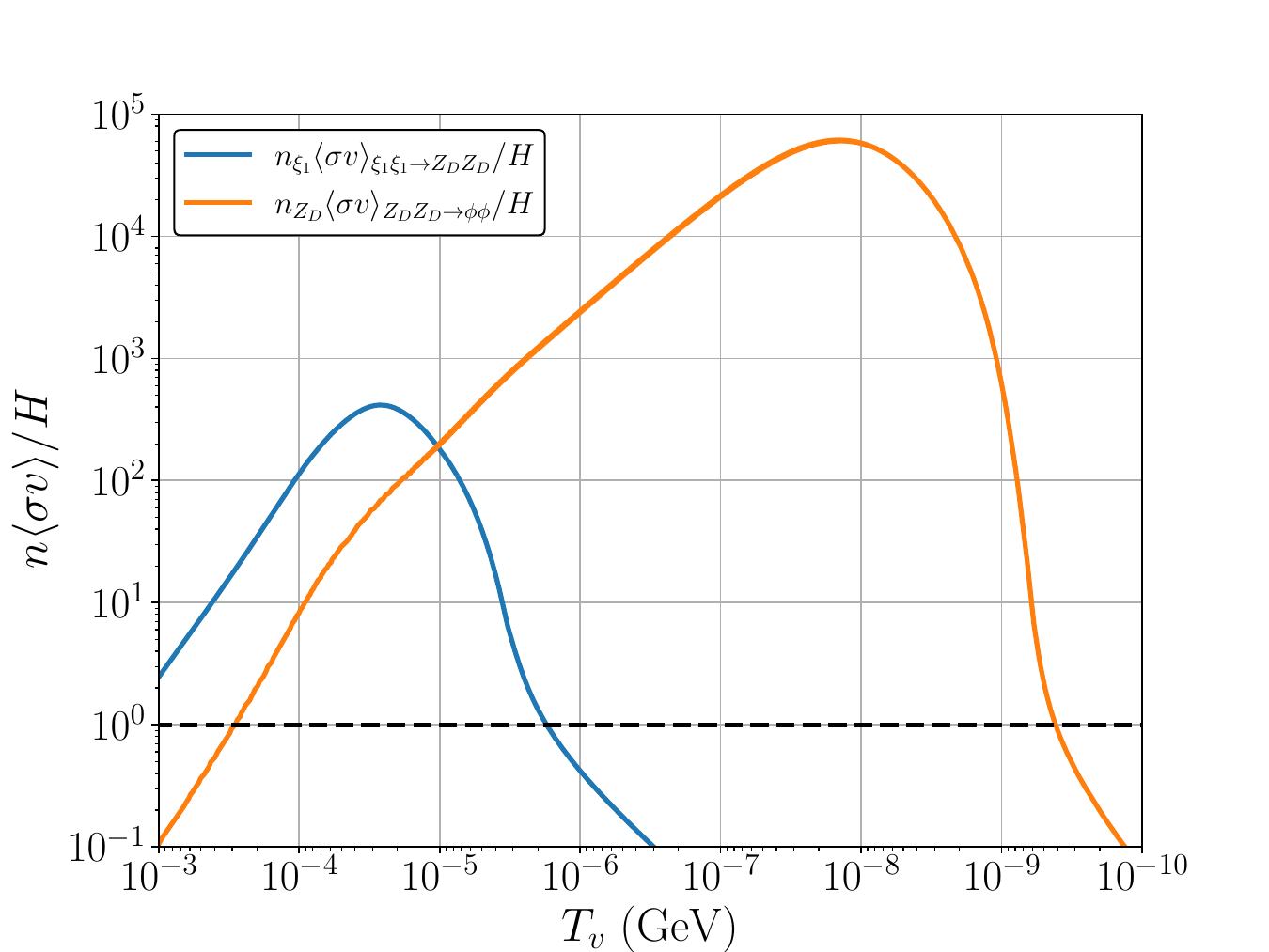}
    \caption{Thermal averaged annihilation cross section between hidden sector particles in terms of a hidden sector temperature $T_h$ (Left), and the ratio of the interaction rate and expansion $n\VEV{\sigma v} / H$ vs $T_v$ (right) to ensure that hidden sector particles are in kinetic equilibrium. Here we use the benchmark values of the parameters as given in \eq{eq:benchmark}.}
    \label{fig:sigma-ds}
\end{figure}

It is important to note that while such a small kinetic mixing parameter can safely evade existing constraints, it is insufficient to generate the required energy density for DR to alleviate the $H_0$ tension. Although it can produce the correct DM relic density for certain DM masses, an additional mechanism is needed to address the DR deficit. To overcome this limitation, we invoke a non-thermal contribution through the decay of an additional singlet scalar, $S$. The late-time decay of $S$ serves a dual purpose: 1) it contributes to achieving the correct DM relic density giving additional freedoms for the other model parameters, and 2) it generates the required contribution of SIDR to $\Delta N_{\rm eff}$. 
% This hybrid production mechanism, combining freeze-in and non-thermal contributions, \cred{allows our model to simultaneously address multiple cosmological challenges while remaining consistent with observational constraints.}

In our analysis, we consider the distinct thermal environments of the hidden and visible sectors, acknowledging that they typically exist at different temperatures. To accurately model the evolution of particle species in both sectors, we employ a two-temperature framework in our Boltzmann equations. This approach allows us to track the number densities of particles, while also accounting for the correlations that arise from the interactions between the two sectors.
Using the energy conservation equations,
\begin{equation}
    \begin{split}
        & \frac{d\rho_v}{dt} + 3H (\rho_v + p_v) = -j_h, \\
        & \frac{d\rho_h}{dt} + 3H (\rho_h + p_h) = j_h,
    \end{split}
\end{equation}
where $\rho_i, p_i$ represent the energy density and pressure density of each sector with $i=v,h$ and $j_h$ is the source term of hidden sector which is defined below. Using the relation between time $t$ and the visible sector temperature $T_v$,
\begin{equation}
    \frac{dT_v}{dt} = -\frac{3H(\rho_v+p_v) + j_h}{{d\rho_v}/{dT_v}}\,.
\end{equation}
we can get the evolution equation for the energy density of the hidden sector~\cite{Aboubrahim:2022gjb}:
\begin{equation}
    \frac{d\rho_{h}}{dT_v} = \frac{3H(1+\omega_{h}) \rho_{h} -j_{h}}{3H(1+\omega_v)\rho_v + j_h} \frac{d\rho_v}{dT_v}.
    \label{eq:rho_hidden}
\end{equation}
Here $\omega_h = p_h/\rho_h$ and $\omega_v = p_v / \rho_v$ refer to the equation of state parameter for the hidden sector and visible sector respectively.
In \eq{eq:rho_hidden}, the source for the hidden sector $j_h$ is given by

\begin{equation}
\begin{split}
    j_h & = n_S m_S \Gamma_S +  \sum_f j(f\bar{f} \to \xi_i \bar{\xi_j})(T_v) + j(f\bar{f} \to Z_D \gamma)(T_v) + j(f\gamma \to f Z_D)(T_v).
\end{split}
\label{eq:sourceterm}
\end{equation}
where the $j$-terms in the above equation are given in Appendix~\ref{app:sourceterm}. In \eq{eq:sourceterm}, the energy density injection into the hidden sector has two distinct contributions: the first term represents the non-thermal component arising from the decay of the scalar S, while the subsequent terms account for the freeze-in production mechanism, where particles from the visible sector gradually populate the hidden sector through feeble interactions.

To conduct our numerical analysis, we have implemented the model in the \texttt{LanHEP} package~\cite{Semenov:2014rea}, which automates the process of generating Feynman rules from the Lagrangian. Subsequently, we utilized \texttt{CalcHEP}~\cite{Belyaev:2012qa} to compute the necessary cross-sections and decay widths. These calculated quantities were then incorporated into a custom-built solver using the \texttt{Boost C++} libraries~\cite{BoostLibrary} to numerically integrate the relevant Boltzmann equations.

Fig.~\ref{fig:sigma-ds} illustrates the temperature dependence of interactions within the hidden sector. The left panel shows the thermally averaged cross-section $\langle\sigma v\rangle$ for various processes as a function of the hidden sector temperature $T_h$, while the right panel depicts the ratio $n\VEV{\sigma v}/H$ for DM annihilations and $Z_D$ annihilations versus the visible sector temperature $T_v$. The thermal averaged cross-sections for $\xi_i$ annihilation to $Z_D$ and $Z_D$ annihilation to DR exhibit distinct temperature regimes: at high temperatures ($T_h \gg M_\xi$), $\langle\sigma v\rangle \propto T_h^2$, and at low temperatures, $\langle\sigma v\rangle$ approaches a constant value. This behavior is characteristic of s-wave annihilation processes involving vector mediators. For $M_{Z_D} \ll M_{\xi_1}$, the cross-section asymptotes to $\langle\sigma v\rangle \approx \pi \alpha^2_D/M^2_{\xi_1}$. As evident from the right panel figure, the condition $n\langle\sigma v\rangle/H \gg 1$ is satisfied across a wide temperature range, ensuring equilibrium among hidden sector particles.

\begin{figure}[tbp]
    \centering
    \includegraphics[width=0.48\textwidth]{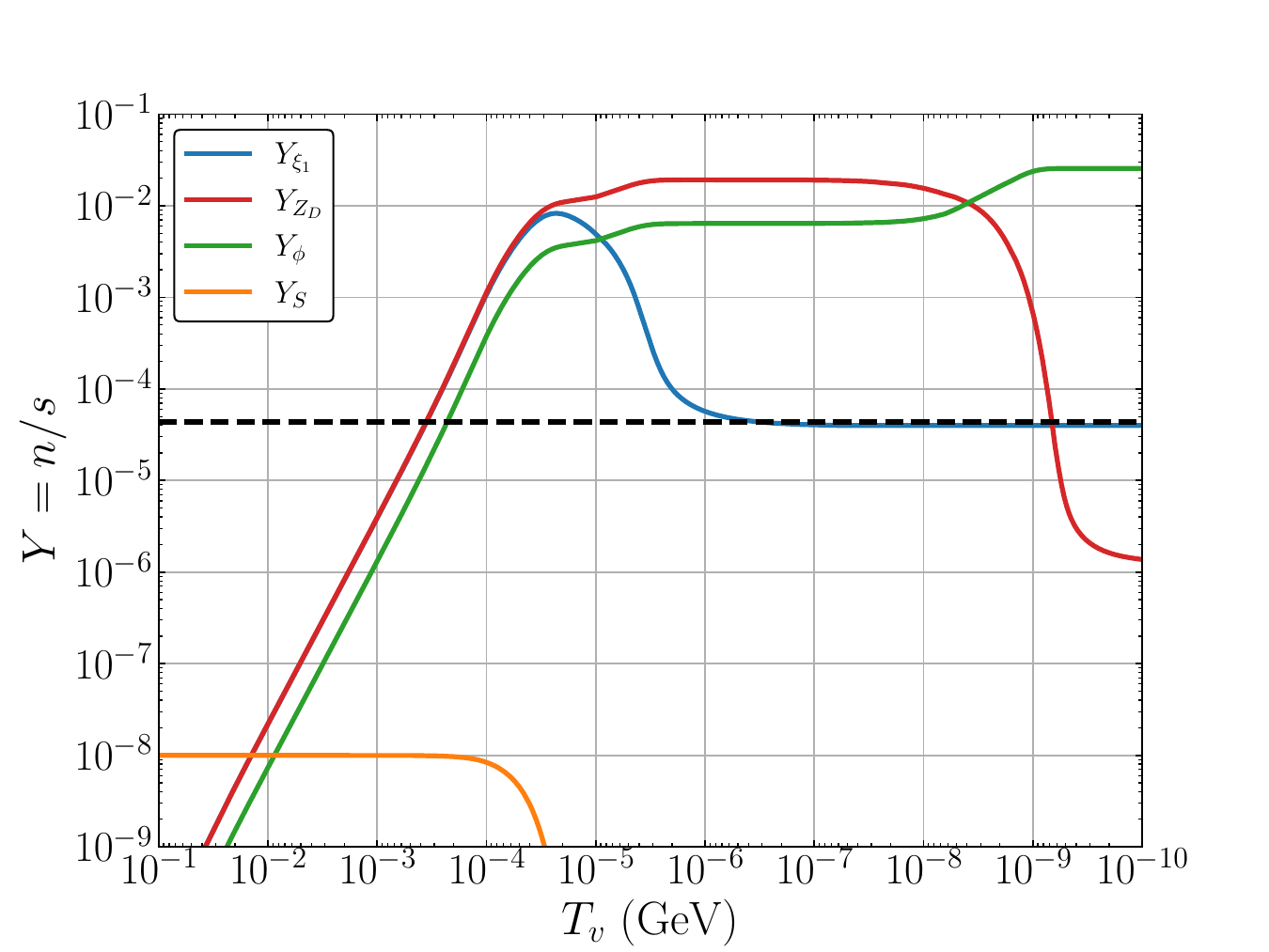}
    \includegraphics[width=0.48\textwidth]{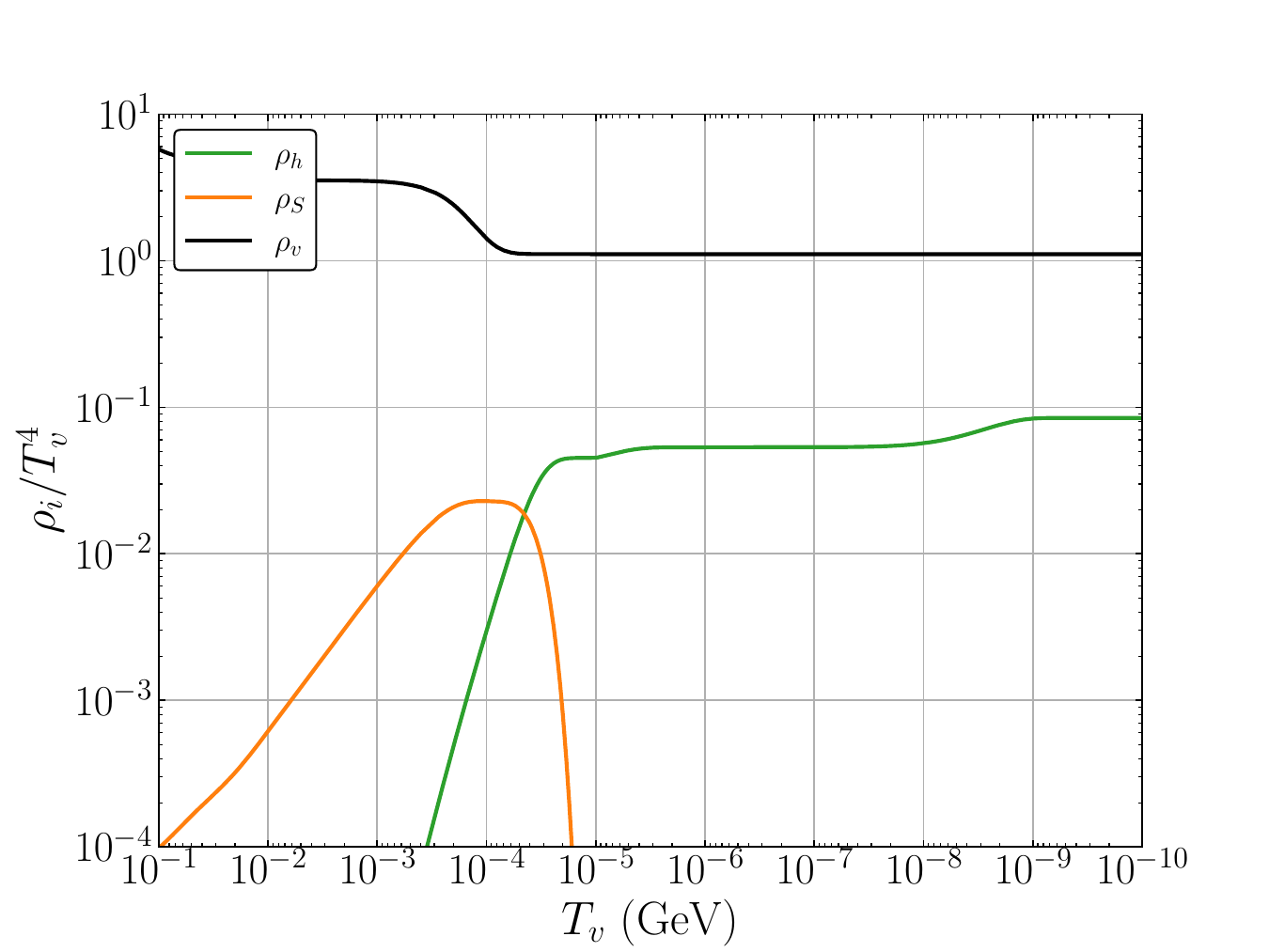}
    \caption{Evolution of the abundance of the hidden particles (left) and the energy density (right). The same parameters are used as in Fig.~\ref{fig:sigma-ds} with $M_S = 130\gev$ and $\Gamma_S=10^{-27}\gev$. }
    \label{fig:Y_rho}
\end{figure}

In Fig.~\ref{fig:Y_rho}, we illustrate the abundance (left) and energy density (right) evolution of hidden sector particles, using the same parameters as in Fig.~\ref{fig:sigma-ds}, with $M_S = 130\ \text{GeV}$ and a decay width $\Gamma_S = 10^{-27}\ \text{GeV}$. 
In the left panel, the blue line shows the number density of dark matter particle $\xi_1$. Initially, it grows due to the decay of $S$. This growth continues until $\xi_1$ annihilates into $Z_D$ around $T_v = 5 \times 10^{-5}\ \rm{GeV}$. After freeze-out, the abundance stabilizes at a value corresponding to the correct relic abundance of DM. The black dashed line represents the comoving number density required for the correct relic density for $M_{\rm DM}=10$ keV. It is important to note that although both $\xi_1$ and $\xi_2$ are produced from the decay of $S$, $\xi_2$ rapidly decays into $\xi_1$ and $\phi$, converting its entire number density to $\xi_1$.
 Following the decay of $S$ and DM annihilation, the total energy density in the hidden sector is dominated by $Z_D$ until $Z_D$ itself freezes out at approximately $10 \unitev$. The number density evolution of $Z_D$ is depicted by the red line. Subsequently, $\rho_h$ is primarily dominated by $\phi$. Notably, we observe a step-like increase in the abundance of $\phi$  after each annihilation process {\it i.e.} DM annihilation into $Z_D$ and $Z_D$ annihilation into $\phi$, as shown by the green line. In the right panel, we compare the energy densities of $S$ (orange), $\rho_h$ (green), and $\rho_v$ (black). As mentioned earlier, the hidden sector energy density $\rho_h$ also exhibits distinct step-like increases following the annihilation events of $\xi_1$ and $Z_D$. This key feature of cascading energy transfer from heavier to lighter species within the dark sector is similar to that in WZDR model~\cite{Aloni:2021eaq}.

\begin{figure}[tbp]
    \centering
    \includegraphics[width=0.48\textwidth]{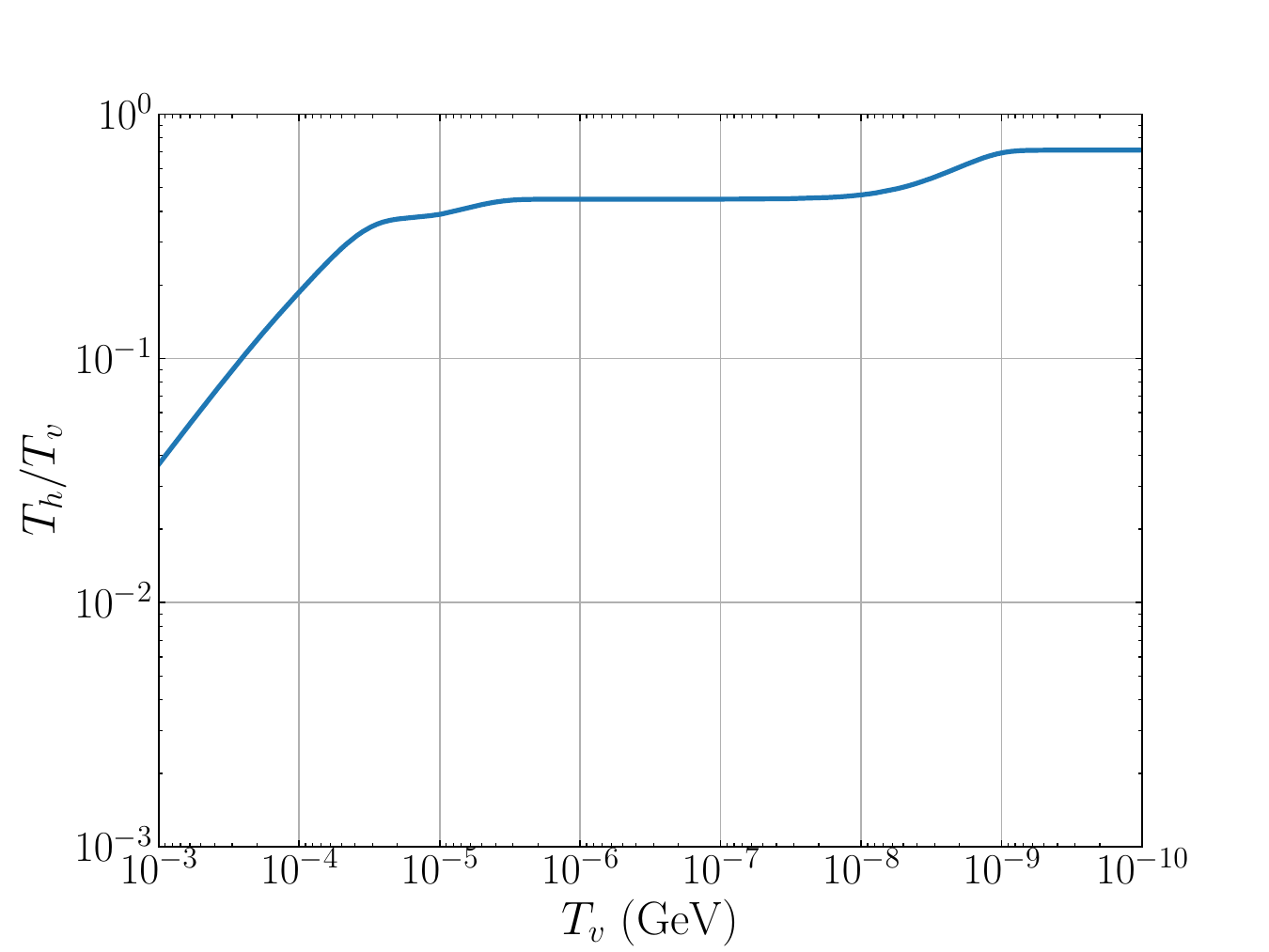}
\includegraphics[width=0.48\textwidth]{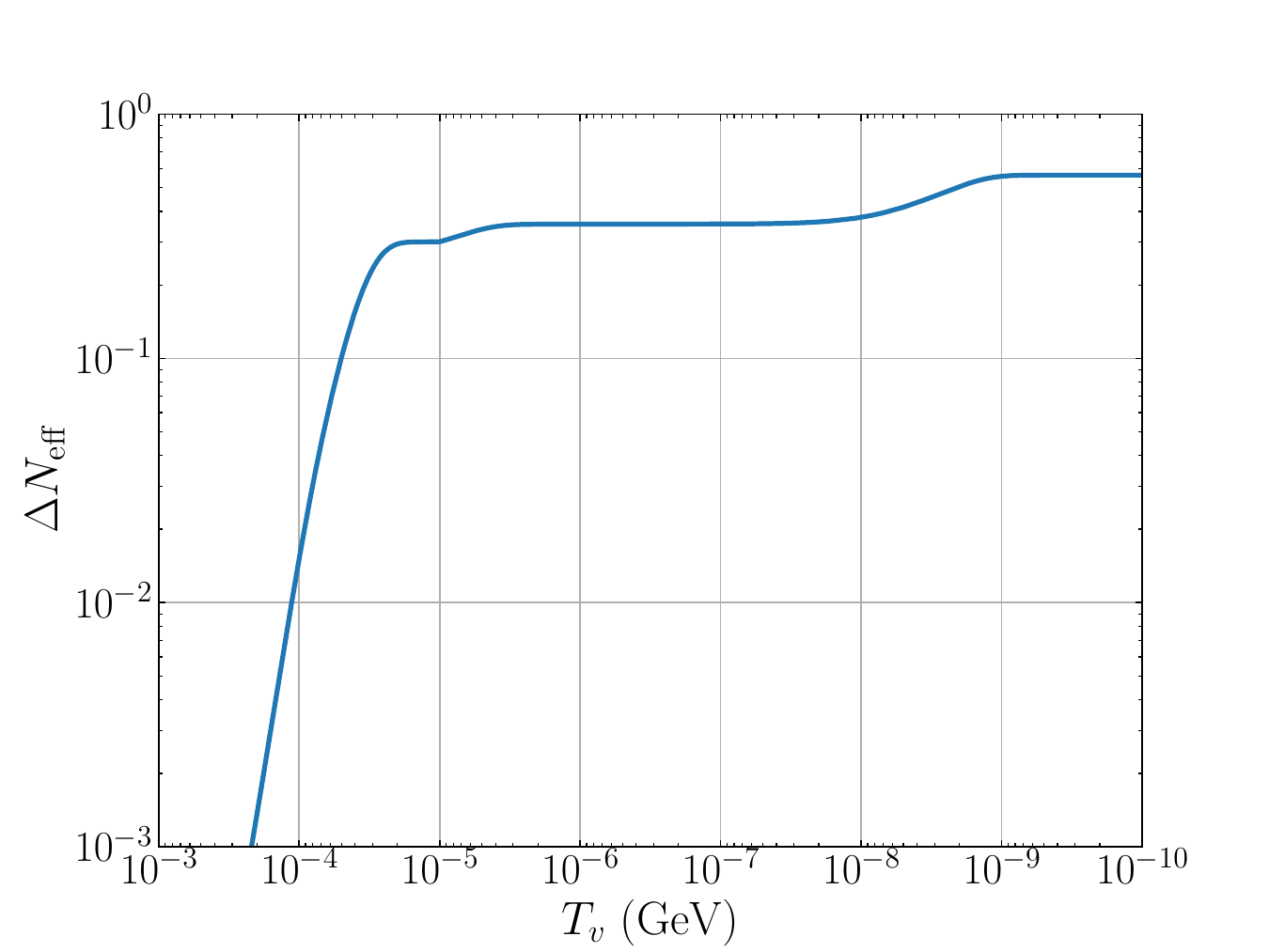}
    \caption{(Left) Evolution of the ratio of hidden sector temperature to the visisble sector temperature ($T_h/T_v$). (Right) The evolution of $\Delta N_{\rm eff}$. The same parameters are used as in Fig.~\ref{fig:sigma-ds} and \ref{fig:Y_rho}.  }
    \label{fig:T_hidden}
\end{figure}

Due to the substantial gauge and Yukawa couplings, the hidden sector particles achieve thermal equilibrium efficiently. We define the hidden sector temperature, $T_h$ from the hidden sector energy density using the relation:
\dis{
\rho_h =\sum_i g_i \frac{\pi^2}{30}T_h^4.
 }
After the decay of $S$, the particles $\xi_i$ and $Z_D$ annihilate into $\phi$, making the hidden sector predominantly composed of dark radiation $\phi$. Thus, the hidden sector energy density can be expressed as:
\begin{equation}
    \rho_h  =  \frac{\pi^2}{30}g_\phi T_h^4,
\end{equation}
In Fig.~\ref{fig:T_hidden}, we depict the evolution of the hidden sector temperature relative to the visible sector temperature, $T_h/T_v$ (left), and the corresponding $\Delta N_{\rm eff}$ (right). We can see that, following the decay of $S$, there is a sharp rise in $T_h$ at $T_v \sim 10^{-5}\ \text{GeV}$. The hidden sector temperature also increases when relatively massive hidden sector particles annihilate into lighter hidden sector particles which is evident around $T_v\sim 10^{-5}\gev$ and $T_v\sim10^{-8}\gev$ respectively. Similarly, $\Delta N_{\rm eff}$ also increases during these annihilation events. 
%\cred{This stepwise increase in the $\Delta N_{\rm eff}$, ultimately helps in enhancing the dark radiation component and its impact on cosmological observables.}

%%%%%%%%%%%%%%%%%%%%%%%%%%%%%%%
\section{Conclusion}\label{sec:conclude}
%%%%%%%%%%%%%%%%%%%%%%%%%%%%%%%
In this paper, we have presented a novel particle physics framework, termed ‘{\bf SIDR}$\boldsymbol{+z_t}$', which offers a comprehensive approach to addressing multiple cosmological tensions observed in recent measurements. 
Our model, which is based on an inelastic dark matter scenario coupled with self-interacting dark radiation within a $U(1)_D$ gauge symmetry extension of the SM, offers a framework that may help address the Hubble tension, $S_8$ tension, and discrepancies observed in Lyman-$\alpha$ data. In future work, we plan to conduct a detailed quantitative study using CMB analysis codes to further explore the potential of our model in addressing and resolving these cosmological anomalies.
%Our model, based on an inelastic dark matter scenario coupled with self-interacting dark radiation within a $U(1)_D$ gauge symmetry extension of the SM, provides a unified solution to the Hubble tension, $S_8$ tension, and discrepancies in Lyman-$\alpha$ observations.
The key features of our model include:
SIDR that behaves as a fluid, suppressing free-streaming effects and anisotropic stress, which is crucial for addressing the $S_8$ tension and 
inelastic DM interacting with DR, providing a mechanism to suppress the matter power spectrum at small scales, potentially reconciling Lyman-$\alpha$ observations.
A distinct temperature dependence for the DM-DR interaction rate, which has a cut-off at the transition redshift $z_t$ determined by the mass-splitting between inelastic dark fermions, sets our model apart from previous SIDR+ and WZDR+ scenarios. The energy scales of the steps for increase in energy density of the two “stepped" DR fluids, being independent of the MPS suppression scale, provides enhanced flexibility in addressing the cosmological tensions.  
The production mechanism for dark sector particles via freeze-in and non-thermal contributions, allows for significant $N_{\rm eff}$ from SIDR without violating BBN constraints, while simultaneously achieving the correct DM relic abundance.
Our comprehensive analysis, including solving the relevant Boltzmann equations, demonstrates the model's potential to uphold BBN predictions while producing additional contributions to $N_{\rm eff}$ prior to recombination. This framework not only addresses current cosmological tensions but also provides a testable particle physics model that can be probed by future experiments.

\section*{Acknowledgements}\label{acknowledge}
The authors acknowledge Amin Aboubrahim and Mellisa Joseph for helpful discussions.
 The authors also acknowledge the financial support from National Research Foundation (NRF) grant
funded by the Korea government (MEST) NRF-2022R1A2C1005050. Hospitality at APCTP during the focus program ``The Origin and Evolution of the Universe'' is kindly acknowledged.
%%%%%%%%%%%%%%%%%%%%%%%%%%%%%%%%%%%%%%%%%%%%%%%%%%%%%%%%%%%%%%%%%%%%%%%%%%%%%%%%%%
		
\appendix

\section{Neutrino Temperature}
In standard cosmology, the energy density around BBN epoch comes mainly from neutrinos, photon, and electron, which can be described as 
\dis{
\rho & = \rho_\nu + \rho_\gamma +\rho_e \\
 & =  \frac{7 N_\nu}{8} \frac{\pi^2}{30} g_\nu T_\nu^4 + \frac{\pi^2}{30}  g_\gamma T_\gamma^4 +  \frac{\pi^2}{30}g_{\rm eff, e}T_\gamma^4
}
where $g_e^{\rm eff}$ is defined from the relation
\dis{
g_e(T) =& \frac{g_e}{2\pi^2} \int_0^\infty dp \frac{p^2\sqrt{p^2 + m_e^2}}{\exp(\sqrt{p^2+m_e^2}/T) \pm 1} \equiv g_{\rm eff, e} \frac{\pi^2}{30} g_e T_\gamma^4. 
}
Using $\rho = g_{\rm eff} \frac{\pi^2}{30} T_\gamma^4$,
we find the change of the temperature ratio during $e^+e^-$ annihilation and after that as
\dis{
\bfrac{T_\nu}{T_\gamma}^4 = \frac{g_{\rm eff} -g_\gamma-g_{\rm eff, e} g_e }{\frac{7}{8} N_\nu g_\nu}.
}

  \section{Thermal averaged cross section and decay rate}

The thermal averaged annihilation cross section is defined as
\begin{equation}
    \expval{\sigma v}_{ab\to cd}(T) = \frac{1}{8 m_a^2 m_b^2 T K_2(m_a/T)K_2(m_b/T)}\int_{s_0}^\infty ds \sqrt{s} (s-s_0) \sigma(s) K_1(\sqrt{s}/T),
\end{equation}
where $s_0$ is minimum value of the Mandelstam variable $s$ in the annihilation process. 
The thermal average decay rate is defined as 
\begin{equation}
    \expval{\Gamma} = \Gamma \frac{K_1(m_S/T_v)}{K_2(m_S/T_v)} \simeq \Gamma,
\end{equation}
where we used $m_S \gg T_v$ in the last equation.

\section{The source terms from the visible to the hidden sector}
\label{app:sourceterm}
Energy source term from the visible sector to the hidden sector is
\dis{
j= \frac{T}{32\pi^4}\int_{s_0}^{\infty} \sigma(s) s (s-s_0) K_2(\sqrt{s}/T),
}
where $K_2(T)$ is the modified Bessel function of the second kind.

\section{Boltzmann equations}
\label{apppendix:be}

The Boltzmann equations for the hidden sector evolution are
%\cred{Add S-decay and $\xi_2$ decay}
%
\begin{equation}
    \begin{split}
        \frac{d Y_{\xi_1}}{d T_v}=  -\frac{s}{H}K_v & \left[ \expval{\sigma v}_{f\bar{f}\to \xi_1\bar{\xi}_1}(T_v) \left(Y^{eq}_{f}(T_v)\right)^2 + \expval{\sigma v}_{f\bar{f}\to \xi_1\bar{\xi}_2}(T_v) \left(Y^{eq}_{f}(T_v)\right)^2 \right. \\
        & + \frac{1}{s}\langle\Gamma_{Z\to \xi_1\bar{\xi}_1}\rangle(T_v)Y_Z^{eq}+\frac{1}{s}\langle\Gamma_{Z\to \xi_1\bar{\xi}_2}\rangle(T_v)Y_Z^{eq} \\
        &+\frac{1}{s}\langle\Gamma_{S\to \xi_1\bar{\xi}_1}\rangle(T_v)Y_S+\frac{1}{s}\langle\Gamma_{\xi_2 \to \xi_1 \phi}\rangle(T_h)Y_{\xi_2}
        \\
        & + \expval{\sigma v}_{\bar{\xi}_1\xi_2\to \xi_1\bar{\xi}_1}(T_h) \left(Y_{\xi_1}Y_{\xi_2}-\frac{Y^{eq}_{\xi_1}(T_h)Y^{eq}_{\xi_2}(T_h)}{Y^{eq}_{\xi_1}(T_h)^2} Y^2_{\xi_1}\right)\\
        & + \expval{\sigma v}_{\xi_2\bar{\xi}_2\to \xi_1\bar{\xi}_1}(T_h) \left(Y^2_{\xi_2}-\left(\frac{Y^{eq}_{\xi_2}(T_h)}{Y^{eq}_{\xi_1}(T_h)}\right)^2 Y^2_{\xi_1}\right) \\
        & -\expval{\sigma v}_{\xi_1\bar{\xi}_1 \to Z_D  Z_D}(T_h)\left(Y^2_{\xi_1}-\left(\frac{Y^{eq}_{\xi_1}(T_h)}{Y^{eq}_{Z_D}(T_h)}\right)^2 Y^2_{Z_D}\right) \\
        & \left. -\expval{\sigma v}_{\xi_1\bar{\xi}_2 \to Z_D  Z_D}(T_h)\left(Y_{\xi_1}Y_{\xi_2}-\frac{Y^{eq}_{\xi_1}(T_h)Y^{eq}_{\xi_2}(T_h)}{Y^{eq}_{Z_D}(T_h)^2} Y^2_{Z_D}\right) \right],
    \end{split}
    \label{eq:BE_xi1}
\end{equation}
\begin{equation}
    \begin{split}
        \frac{d Y_{\xi_2}}{d T_v}=  -\frac{s}{H}K_v  & \left[\expval{\sigma v}_{f\bar{f}\to \xi_2\bar{\xi}_2}(T_v) \left(Y^{eq}_{f}(T_v)\right)^2+\expval{\sigma v}_{f\bar{f}\to \xi_1\bar{\xi}_2}(T_v) \left(Y^{eq}_{f}(T_v)\right)^2 \right. \\
        & + \frac{1}{s}\langle\Gamma_{Z\to \xi_2\bar{\xi}_2}\rangle(T_v)Y_Z+\frac{1}{s}\langle\Gamma_{Z\to \xi_1\bar{\xi}_2}\rangle(T_v)Y_Z \\
        &+\frac{1}{s}\langle\Gamma_{S\to \xi_2\bar{\xi}_2}\rangle(T_v)Y_S-\frac{1}{s}\langle\Gamma_{\xi_2 \to \xi_1 \phi}\rangle(T_h)Y_{\xi_2}\\
        & - \expval{\sigma v}_{\bar{\xi_1}\xi_2\to \xi_1\bar{\xi}_1}(T_h) \left(Y_{\xi_1}Y_{\xi_2}-\frac{Y^{eq}_{\xi_1}(T_h)Y^{eq}_{\xi_2}(T_h)}{Y^{eq}_{\xi_1}(T_h)^2} Y^2_{\xi_1}\right) \\
        & -\expval{\sigma v}_{\xi_2\bar{\xi}_2\to \xi_1\bar{\xi}_1}(T_h) \left(Y^2_{\xi_2}-\left(\frac{Y^{eq}_{\xi_2}(T_h)}{Y^{eq}_{\xi_1}(T_h)}\right)^2 Y^2_{\xi_1}\right) \\
        & -\expval{\sigma v}_{\xi_2\bar{\xi}_2 \to Z_D  Z_D}(T_h)\left(Y^2_{\xi_2}-\left(\frac{Y^{eq}_{\xi_2}(T_h)}{Y^{eq}_{Z_D}(T_h)}\right)^2 Y^2_{Z_D}\right) \\
        & \left. -\expval{\sigma v}_{\bar{\xi}_1\xi_2 \to Z_D  Z_D}(T_h)\left(Y_{\xi_1}Y_{\xi_2}-\frac{Y^{eq}_{\xi_1}(T_h)Y^{eq}_{\xi_2}(T_h)}{Y^{eq}_{Z_D}(T_h)^2} Y^2_{Z_D}\right) \right],
    \end{split}
    \label{eq:BE_xi2}
\end{equation}
\begin{equation}
\begin{split}
    \frac{d Y_{Z_D}}{d T_v}= -\frac{s}{H}K_v & \left[ \langle\sigma v \rangle_{f \bar{f} \to Z_D  \gamma}(T_v)\left(Y^{eq}_{f}(T_v)\right)^2 + 2\langle\sigma v \rangle_{f \gamma \to f Z_D }(T_v)\left(Y^{eq}_{f}(T_v)\right)^2 \right. \\
    &\left. +  2\expval{\sigma v}_{\xi_1\bar{\xi_1} \to Z_D  Z_D}(T_h)\left(Y_{\xi_1}^2-\left(\frac{Y^{eq}_{\xi_1}(T_h)}{Y^{eq}_{Z_D}(T_h)}\right)^2 Y^2_{Z_D}\right)  \right. \\
    & + 2 \expval{\sigma v}_{\xi_2\bar{\xi_2} \to Z_D  Z_D}(T_h)\left(Y_{\xi_2}^2-\left(\frac{Y^{eq}_{\xi_2}(T_h)}{Y^{eq}_{Z_D}(T_h)}\right)^2 Y^2_{Z_D}\right)  \\
    & + 2 \expval{\sigma v}_{\xi_1\bar{\xi_2} \to Z_D  Z_D}(T_h)\left(Y_{\xi_1}Y_{\bar{\xi_2}}-\frac{Y^{eq}_{\xi_1}(T_h)Y^{eq}_{\xi_2}(T_h)}{Y^{eq}_{Z_D}(T_h)^2} Y^2_{Z_D}\right)  \\
    & + 2 \expval{\sigma v}_{\xi_2\bar{\xi_1} \to Z_D  Z_D}(T_h)\left(Y_{\bar{\xi_1}}Y_{\xi_2}-\frac{Y^{eq}_{\xi_1}(T_h)Y^{eq}_{\xi_2}(T_h)}{Y^{eq}_{Z_D}(T_h)^2} Y^2_{Z_D}\right)  \\
    & \left. - 2\langle\sigma v\rangle_{Z_D Z_D \rightarrow \phi\phi} \left(Y_{Z_D}^2- \left(\frac{Y^{eq}_{Z_D}(T_h)}{Y^{eq}_{\phi}(T_h)}\right)^2Y_\phi^2\right)\right],
\end{split}
\label{eq:BE_Zx}
\end{equation}
\begin{equation}
    \frac{dY_\phi}{dT_v} = -\frac{s}{H} K_v \left[ 2\langle\sigma v\rangle_{Z_D Z_D \rightarrow \phi\phi} \left(Y_{Z_D}^2- \left(\frac{Y^{eq}_{Z_D}(T_h)}{Y^{eq}_{\phi}(T_h)}\right)^2Y_\phi^2\right)\right],
\end{equation}
\begin{equation}
    \frac{dY_S}{dT_v} = -\frac{s}{H} K_v \frac{1}{s} \expval{\Gamma(S\to \xi_i\bar{\xi_i})} Y_S
\end{equation}
where $K_v$ is defined by
\begin{equation}
    K_v \equiv \frac{d \rho_v / dT_v}{4\zeta\rho_v + 4(\zeta -\zeta_h) \rho_h + j_h/H}\sim \frac{1}{T_v}.
    \label{Kv}
\end{equation}
where $\zeta= \frac{p_v + p_h}{\rho_v + \rho_h}$, $\zeta_h = \frac{3}{4} (1+\frac{p_h}{\rho_h})$ and $j_h$ is the source term for the hidden sector. The last approximation in \eq{Kv} is applied when the source term is subdominant to the energy density in the visible sector.

\bibliographystyle{JHEP}
\bibliography{ref}
		
\end{document}